\documentclass[%
 reprint,
 amsmath,amssymb,
 aps,
]{revtex4-2}

\usepackage{graphicx}
\usepackage{dcolumn}
\usepackage{bm}
\usepackage{physics}
\usepackage{xcolor}
\usepackage{makecell}

\renewcommand{\tensor}[1]{\overline{\overline{{#1}}}}

\newcommand{\sgn}[1]{~\mathrm{sgn}(#1)}
\newcommand{\hc}{\mathrm{H.c.}}

\begin{document}

\preprint{APS/123-QED}

\title{Polariton-Assisted Resonance Energy Transfer Beyond Resonant Dipole-Dipole Interaction: A Transition Current Density Approach}

\author{Ming-Wei Lee}
\affiliation{%
Department of Physics, National Taiwan University, Taipei 10617, Taiwan
}%
\affiliation{Institute of Atomic and Molecular Sciences, Academia Sinica, Taipei 10617, Taiwan}

\author{Liang-Yan Hsu}%
\email{lyhsu@gate.sinica.edu.tw}
\affiliation{%
Institute of Atomic and Molecular Sciences, Academia Sinica, Taipei 10617, Taiwan
}
\affiliation{%
Department of Chemistry, National Taiwan University, Taipei 10617, Taiwan
}
\affiliation{National Center for Theoretical Sciences, Taipei 10617, Taiwan}%

\date{\today}

\begin{abstract}
Using electric dipoles to describe light-matter interactions between two entities is a conventional approximation in physics, chemistry, and materials science.
However, the lack of material structures makes the approximation inadequate when the size of an entity is comparable to the spatial extent of electromagnetic fields or the distance of two entities.
In this study, we develop a unified theory of radiative and non-radiative resonance energy transfer based on transition current density in a theoretical framework of macroscopic quantum electrodynamics. The proposed theory allows us to describe polariton-assisted resonance energy transfer between two entities with arbitrary material structures in spatially dependent electromagnetic fields.
To demonstrate the generality of the proposed theory, we rigorously prove that our theory can cover the main results of the transition density cube method and the plasmon-coupled resonance energy transfer. 
We believe that this study opens a promising direction for exploring light-matter interactions beyond the scope of electric dipoles and provides new insights into materials physics. 

\end{abstract}

\maketitle


\section{Introduction}
Resonance energy transfer (RET) is a fundamental process in photophysics and has attracted considerable attention in a variety of fields owing to its extensive applications in biological and chemical sensing \cite{Wu2020,Chen2013,Chang1995,Zadran2012,Kaminska2021,Shi2015,Xia2009,Hildebrandt2017,Bhuckory2019}, molecular imaging~\cite{Li2020,Sasmal2016,Ha2001,Zhou2019,Choi2009}, and photovoltaics~\cite{Scully2007,Liu2005,Shankar2009,Li2015,Heidel2007}.
To understand the mechanism of RET, numerous theoretical studies have been conducted based on the electric-dipole approximation (EDA) since 1940s~\cite{Dexter1953,Skourtis2016,Murphy2004,Olaya-Castro2011,Duque2015,Weeraddana2016,Salam2018,Salam2022,Andrews2022}.
For example, F\"orster developed a concept of spectral overlap for the description of RET~\cite{Forster1948,Forster1949}, Andrews established a unified theory of radiative and nonradiative RET~\cite{Andrews1989,Andrews2004,Daniels2003} (which corresponds to the resonant dipole-dipole interaction in free space) from molecular quantum electrodynamics (QED)~\cite{Craig1998}, and Welsch \textit{et al.} incorporated the effect of dielectric environment into the resonant dipole-dipole interaction (RDDI)~\cite{dung2002,dung2002intermolecular}.
These theories not only advance the understanding of RET but also successfully capture the main features of RET.

Recently, experimental studies have shown that the RET rates between two entities (molecules~\cite{Komarala2008,L.-Viger2011,Georgiou2021,Andrew2004,Hichri2021}, semiconductors~\cite{Li2015}, biomolecules~\cite{Zhang2007,Stobiecka2015}, etc.) can be significantly influenced by the presence of polaritons, which provides a new perspective on exploring light-matter interaction in spatially
dependent electromagnetic fields. It is well-known that the EDA cannot be used to describe RET in the following scenarios, (i) the distance between two entities is insufficient~\cite{Munoz-Losa2009,Andrews2011,Beljonne2009,Zheng2014,Andrews2009}, and (ii) the surrounding electromagnetic field of entities varies drastically~\cite{Zhang2020,NasiriAvanaki2018}.
In the former scenario, several approaches have been successively proposed and applied to some representative systems, e.g., transition monopole theory for chlorophylls~\cite{Chang1999}, line-dipole approximation for conjugated polymers~\cite{Beenken2004}, and transition density cube (TDC) method for pigments of light-harvesting complex~\cite{Krueger1998}, but these approaches cannot describe the retardation effect, i.e., the mechanism of radiative RET.
In the later scenario, a straightforward improvement is to consider the quadrupolar interaction~\cite{Scholes1997,NasiriAvanaki2018,Salam2005}, but the convergence of multipolar expansion depends on material structures. 
Moreover, the methods used to address the scenarios (i) and (ii) cannot describe the influence of polaritons (photons dressed by dielectric environments) on RET. Therefore, to address the above issues, our strategy is to start from the framework of macroscopic QED \cite{Vogel2006,Buhmann2012} (which enables us to incorporate the effect of dielectric environments) and derive an explicit RET-rate expression by employing a transition current density approach (which enables us to capture the retardation effects).



In this article, the main goal is to establish a generalized theory of RET beyond the EDA and allow us to study RET between two entities with material structures in spatially dependent vacuum electric fields.
The structure of this article is organized as follows.
In Sec.~\ref{sect:theory}, we begin from the total Hamiltonian of polaritons and point charges with interactions introduced via the minimal coupling procedure.
Next, we derive the RET rate of a pair of molecules expressed by the molecular transition current density by expanding Born series to the second order.
Moreover, we adopt the Condon-like approximation in order to separate the electronic and nuclear degrees of freedom, and then derive a formula which allows us to acquire the transition current density via \textit{ab initio} calculations.
In Sec.~\ref{sect:app}, we demonstrate the generality of our theory by comparing them to previous works, including the TDC method~\cite{Krueger1998} and plasmon-coupled RET \cite{Hsu2017}.
In the last section, we present a brief summary of this research.

\section{Theory}
\label{sect:theory}
\subsection{Total Hamiltonian}
In the non-relativistic regime, we consider a collection of point charges with the quantized electromagnetic fields in the presence of linear, dispersive, and absorbing media (polaritons).
Note that the formation of polaritons originates from the hybridization of photons and media, which does not include the contribution of point charges.
Hence, in this context, the total Hamiltonian comprises a polariton Hamiltonian and a Hamiltonian of point charges (pc),
$\hat{H}_\mathrm{tot}=\hat{H}_\mathrm{pol}+\hat{H}_\mathrm{pc}$.
To properly describe the quantum behavior of polaritons, we adopt the quantization framework of macroscopic QED~\cite{dung2002intermolecular,Vogel2006,Buhmann2012} and express the polariton Hamiltonian as
\begin{align}
    \hat{H}_\mathrm{pol} = \int \dd[3]{\vb{r}} \int_0^\infty \dd{\omega}\hbar\omega \hat{\mathbf{f}}^\dagger(\mathbf{r},\omega)\cdot\hat{\mathbf{f}}(\mathbf{r},\omega),
\end{align}
where the vector bosonic operators, $\hat{\mathbf{f}}^\dagger(\mathbf{r},\omega)$ and $\hat{\mathbf{f}}(\mathbf{r},\omega)$, are the creation and annihilation operators that obey the following commutation relations,
\begin{subequations}
    \begin{align}
        \left[\hat{f}_k^{\mathstrut{}}(\mathbf{r},\omega),\hat{f}_{k'}^\dagger(\mathbf{r'},\omega')\right] &= \delta_{kk'}\delta(\mathbf{r-r'})\delta(\omega-\omega'),\\
        \left[\hat{f}_k(\mathbf{r},\omega),\hat{f}_{k'}^{\hphantom{\dagger}}(\mathbf{r'},\omega')\right] &= 0.
    \end{align}
    \label{Eq:commutation}%
\end{subequations}
Incidentally, the macroscopic QED can be reduced to the microscopic (or molecular) QED \cite{Craig1998,Salam2009} in the cases of homogeneous dilute media or vacuum \cite{Gruner1996}.
For the Hamiltonian of point charges, the interaction between point charges and quantized electromagnetic fields is introduced through the minimal coupling procedure in the Coulomb gauge,
\begin{align}
    \nonumber
    \hat{H}_\mathrm{pc} &= 
    \sum_{n} \frac{1}{2m_n}\left[\vphantom{\frac{1}{2m_n}}\hat{\mathbf{p}}_n - q_n \hat{\mathbf{A}}(\hat{\mathbf{r}}_n) \right]^2
    +\sum_{n<m} \hat{V}(\hat{\vb{r}}_n,\hat{\vb{r}}_m)\\
    &~+\sum_{n} q_n\hat{\varphi}(\hat{\vb{r}}_n).
\end{align}
The first term in $\hat{H}_\mathrm{pc}$ describes the mechanical kinetic energy of point charges in the presence of electromagnetic fields, where $m_n$, $q_n$, $\hat{\vb{r}}_n$, and $\hat{\vb{p}}_n$ are the mass, charge number, position operator, and canonical momentum operator of the $n$-th point charge, respectively \cite{ScullyQuantumOptics}.
The second term, $\hat{V}(\hat{\vb{r}}_n,\hat{\vb{r}}_m)$, describes the Coulomb interaction between the $n$-th and $m$-th point charges.
The final term is the interaction between point charges and scalar potential from media.
In the Coulomb gauge, the scalar potential operator $\hat{\varphi}(\vb{r})$  is associated with the longitudinal auxiliary electric-field operator $\hat{\mathbf{E}}^\parallel (\mathbf{r},\omega)$ [Eq.~(\ref{Eq:defgradphi})].
Similarly, the vector potential operator $\hat{\vb{A}}(\vb{r})$ is associated with the transverse auxiliary electric-field operator $\hat{\mathbf{E}}^\perp (\mathbf{r},\omega)$ [Eq.~(\ref{Eq:defA})].
\begin{subequations}
    \begin{align}
        \label{Eq:defgradphi}
        -\nabla \hat{\varphi}(\vb{r}) &= \int_0^\infty \dd{\omega}
        \left[\hat{\mathbf{E}}^\parallel (\mathbf{r},\omega) + \mathrm{H.c.}\right] 
        ,\\
        \label{Eq:defA}
        \hat{\mathbf{A}}(\mathbf{r}) &= \int_0^\infty \dd{\omega}
        \left[(i\omega)^{-1}\hat{\mathbf{E}}^\perp(\mathbf{r},\omega) + \mathrm{H.c.}\right].
    \end{align}
    \label{Eq:FieldsinFeq}%
\end{subequations}
Here, the longitudinal (transverse) component is defined by the auxiliary electric-field operator $\hat{\mathbf{E}} (\mathbf{r},\omega)$,
\begin{align}
     \hat{\mathbf{E}}^{\parallel(\perp)}(\mathbf{r},\omega)= \int \dd[3]{\vb{r}'} \tensor{\boldsymbol{\delta}}\vphantom{\delta}^{\parallel(\perp)}(\vb{r-r}')\cdot\hat{\mathbf{E}}(\mathbf{r}',\omega),
\end{align}
where $\tensor{\boldsymbol{\delta}}\vphantom{\delta}^{\parallel}(\vb{r-r}')$ and $\tensor{\boldsymbol{\delta}}\vphantom{\delta}^{\perp}(\vb{r-r}')$ denote the longitudinal and transverse dyadic delta functions, respectively.
Moreover, the auxiliary electric-field operator in the frequency domain is defined by \cite{Dung1998,dung2002intermolecular}
\begin{align}
    \nonumber
    &\hat{\mathbf{E}}(\mathbf{r},\omega)\\
    &=i\sqrt{\frac{\hbar}{\pi\epsilon_0}}\frac{\omega^2}{c^2} \int \dd[3]{\mathbf{r'}} \sqrt{\epsilon_\mathrm{I}(\mathbf{r'},\omega)}~\tensor{\mathbf{G}}(\mathbf{r},\mathbf{r'},\omega)\cdot\hat{\mathbf{f}}(\mathbf{r'},\omega).
    \label{Eq:Edef}
\end{align}
Here, $\epsilon_\mathrm{I}(\vb{r}',\omega)=\mathrm{Im}\left[\epsilon(\vb{r}',\omega)\right]$ denotes the imaginary part of the relative permittivity function, and $\tensor{\mathbf{G}}(\mathbf{r},\mathbf{r'},\omega)$ is the dyadic Green's function of macroscopic Maxwell's equations, i.e.,
\begin{align}
    \left[
    \frac{\omega^2\epsilon(\vb{r},\omega)}{c^2}-\nabla\times\nabla\times
    \right]
    \tensor{\vb{G}}(\vb{r},\vb{r}',\omega)=-\tensor{\vb{I}}\delta(\vb{r}-\vb{r}'),
\end{align}
where $\tensor{\vb{I}}$ is a 3-by-3 identity matrix.

On the basis of the total Hamiltonian, we further categorize the point charges according to their belonging entities because the main purpose of this study is to focus on the RET between the entities, where the entities can be atoms, molecules, 2D materials, etc.
Accordingly, the Hamiltonian of point charges is rewritten as
\begin{align}
    \nonumber
    \hat{H}_\mathrm{pc}&=
    \sum_M\hat{H}_M + \sum_{M<M'}\hat{V}_{MM'}
    +\sum_M\left[\hat{V}_{\mathrm{pol},M}+\hat{\Lambda}_{\mathrm{pol},M}\right],
\end{align}
where $M$ denotes the index of entities and $\xi\in M$ denote the index of point charges (including nuclei and electrons) in the entity $M$.
Here, the Hamiltonian $\hat{H}_M$ of the entity $M$ and the Coulomb interaction between two entities $\hat{V}_{MM'}$ are defined by
\begin{align}
    \hat{H}_M &= 
    \sum_{\xi\in M}\frac{\hat{\mathbf{p}}_\xi^2}{2m_\xi}
    +\sum_{\xi<\zeta} \hat{V}(\hat{\vb{r}}_\xi,\hat{\vb{r}}_{\zeta}),
    \label{Eq:HMdef}\\
    \hat{V}_{MM'} &= 
    \sum_{\xi\in M}\sum_{\zeta\in M'}
    \hat{V}(\hat{\vb{r}}_\xi,\hat{\vb{r}}_{\zeta}),
\end{align}
respectively.
The interactions between polaritons and the entity $M$ ($\hat{V}_{\mathrm{pol},M}$ and $\hat{\Lambda}_{\mathrm{pol},M}$) are defined as

\begin{subequations}
\begin{align}
    \hat{V}_{\mathrm{pol},M}&=
    \sum_{\xi\in M} \Big[q_\xi\hat{\varphi}(\hat{\vb{r}}_\xi) - \frac{q_\xi}{m_\xi}\hat{\vb{A}}(\hat{\vb{r}}_\xi)\cdot\hat{\vb{p}}_\xi\Big],
    \label{Eq:Vpolproto}\\
    \hat{\Lambda}_{\mathrm{pol},M}&=\sum_{\xi\in M}\frac{q_\xi^2}{2m_\xi}
    \hat{\vb{A}}^2(\hat{\vb{r}}_\xi).
\end{align}
\label{Eq:fullcoupling}
\end{subequations}

\noindent
Note that in Eq. (\ref{Eq:Vpolproto}), we use the relation, $\hat{\vb{p}}_\xi\cdot\hat{\vb{A}}(\hat{\vb{r}}_\xi)=\hat{\vb{A}}(\hat{\vb{r}}_\xi)\cdot\hat{\vb{p}}_\xi$, which holds as we choose the Coulomb gauge.
In addition, it is convenient to define the charge density operator ${\hat{\rho}}_M(\vb{r})$ and the current density operator ${\hat{\vb{j}}}_M(\vb{r})$ of the entity $M$,
\begin{align}
    \label{Eq:Mrhodef}
    {\hat{\rho}}_M(\vb{r}) &\equiv \sum_{\xi\in M} q_\xi
    \delta(\mathbf{r}-\hat{\mathbf{r}}_\xi),\\
    {\hat{\vb{j}}}_M(\vb{r}) &\equiv
    \sum_{\xi\in M}\frac{q_\xi}{2m_\xi}\Big[
    \delta(\vb{r}-\hat{\vb{r}}_\xi)\hat{\vb{p}}_\xi+
    \hat{\vb{p}}_\xi\delta(\vb{r}-\hat{\vb{r}}_\xi)
    \Big],
    \label{Eq:Mcurrentdef}
\end{align}
so that $\hat{V}_{\mathrm{pol},M}$ can be expressed in a continuous form,

\begin{align}
    \hat{V}_{\mathrm{pol},M}
    &=
    \int \dd[3]{\vb{r}}
    \left[\hat{\varphi}(\mathbf{r}){\hat{\rho}}_M(\mathbf{r})-
    \hat{\mathbf{A}}(\mathbf{r})\cdot{\hat{\vb{j}}}_M(\vb{r})\right].
    \label{Eq:Hintdef}
\end{align}
Note that the order of delta functions and canonical momentum operators in Eq.~(\ref{Eq:Mcurrentdef}) should be done with care because $\hat{\vb{r}}_\xi$ and $\hat{\vb{p}}_\xi$ are not commutative. The details can be found in Appendix \ref{sect:deltap}.
In contrast to $\hat{V}_{\mathrm{pol},M}$, which includes both scalar and vector potentials, $\hat{\Lambda}_{\mathrm{pol},M}$ depends only on the vector potential and  is associated with the diamagnetic response \cite{Rossi2017,Aucar1999,Alhambra2014} of the entity $M$.
Here, we would like to emphasize that the interactions in Eqs.~(\ref{Eq:fullcoupling}a) and (\ref{Eq:fullcoupling}b) are the complete couplings between polaritons and molecules, i.e., the couplings beyond the commonly used EDA.
In other words, the information of molecular structures is fully preserved in Eq.~(\ref{Eq:fullcoupling}).
Finally, the total Hamiltonian is reorganized as the following compact form,
\begin{align}
    \nonumber
    \hat{H}_\mathrm{tot}
    & = \hat{H}_\mathrm{pol} + \sum_M\hat{H}_M \\
    &~+\sum_M\left[\hat{V}_{\mathrm{pol},M}+\hat{\Lambda}_{\mathrm{pol},M}\right] 
    + \sum_{M<M'}V_{MM'}.
    \label{Eq:GeneralFormH}
\end{align}
It is worth pointing out that the above Hamiltonian is a general form for entities coupled to the quantized electromagnetic fields in the presence of dispersive and absorbing media, which can also be applied to investigate other photophysical (photochemical) topics such as spontaneous emission and electron transfer under the influence of polaritons.

\subsection{Transfer Rate of a Two-Entity System}
Now, we focus on the RET between a pair of entities with the assistance of polaritons in the incoherent limit \cite{Zimanyi2010,Nazir2009,Olaya-Castro2011,Wang2022,Chuang2022}, i.e., the RET processes can be described by kinetic rates.
According to the definition in Eq.~(\ref{Eq:GeneralFormH}), the total Hamiltonian in the two-entity case reads
\begin{align}
    \hat{\mathcal{H}}&\equiv\hat{H}_\mathrm{tot}(M=\{A,B\})
    = \hat{\mathcal{H}}_0 + \hat{\mathcal{H}}_1,
\end{align}
with the unperturbed Hamiltonian $\hat{\mathcal{H}}_0$ and interaction Hamiltonian $\hat{\mathcal{H}}_1$ described by
\begin{align}
    \hat{\mathcal{H}}_0 &= \hat{H}_\mathrm{pol} + \hat{H}_A + \hat{H}_B,\\
    \nonumber
    \hat{\mathcal{H}}_1 &= \hat{V}_{\mathrm{pol},A} + \hat{\Lambda}_{\mathrm{pol},A}
    + \hat{V}_{\mathrm{pol},B} + \hat{\Lambda}_{\mathrm{pol},B} + \hat{V}_{AB}\\
    &\approx
    \hat{V}_{\mathrm{pol},A} + \hat{V}_{\mathrm{pol},B} + \hat{V}_{AB}.
\end{align}
In the interaction Hamiltonian, we do not consider the diamagnetic effect in the RET (i.e., we neglect the contribution from the diamagnetic terms, $\hat{\Lambda}_{\mathrm{pol},A}$ and $\hat{\Lambda}_{\mathrm{pol},B}$) because their magnitude is far less than that of the remaining terms \cite{Ashcroft1976,Schatz2002}.
In this system, we consider the initial (final) state, which is the direct product of the polaritonic vacuum state and the energy eigenstates $A$ and $B$,
\begin{subequations}
    \begin{align}
        \ket{{i}} &=
        \ket{a'}\otimes\ket{b}\otimes\ket{\left\{0\right\}},
        ~~ E_i=E_{a'}+E_b,\\
        \ket{{f}} &=
        \ket{a}\otimes\ket{b'}\otimes\ket{\left\{0\right\}},
        ~~ E_f=E_a+E_{b'}.
    \end{align}
\end{subequations}
Here, $\ket{a(a')}$ and $\ket{b(b')}$ are the eigenkets of the Hamiltonians $\hat{H}_{A}$ and $\hat{H}_{B}$, and their corresponding energy are $E_{a(a')}$ and $E_{b(b')}$.
Also, we denote the energy of the initial (final) state to $E_i$ ($E_f$), and we require that $E_{a'}>E_a$ and $E_{b'}>E_b$.
By expanding Born series up to the second order in the time-dependent perturbation theory~\cite{CohenTannoudji1998}, the total RET rate $\Gamma$ is expressed as follows,
\begin{align}
    \Gamma = \sum_{f,i} P_i \Gamma_{fi},\quad
    \Gamma_{fi} = \frac{2\pi}{\hbar}\abs{\bra{{f}}\hat{\mathcal{T}}\ket{{i}}}^2\delta(E_{f}-E_{i}),
    \label{Eq:ratedef}
\end{align}
where $P_i$ is the probability of the inital state and the transition operator $\hat{\mathcal{T}}$ is given by
\begin{align}
    \hat{\mathcal{T}} &= \hat{\mathcal{T}}_1 + \hat{\mathcal{T}}_2 
    = 
    \hat{\mathcal{H}}_1 + \hat{\mathcal{H}}_1\hat{\mathcal{G}_0}\hat{\mathcal{H}}_1.
\end{align}
Here, $\hat{\mathcal{G}}_0$ is the retarded Green's operator, which is defined by
\begin{align}
    \hat{\mathcal{G}}_0=\frac{1}{E_{i}-\hat{\mathcal{H}}_0+i\eta},
    \quad \eta\rightarrow{0}^+.
\end{align}
Now, we evaluate the total transition amplitude $\bra{{f}}\hat{\mathcal{T}}\ket{{i}}$, and divide $\bra{{f}}\hat{\mathcal{T}}\ket{{i}}$ into two parts, $\matrixel{f}{\hat{\mathcal{T}}_1}{i}$ and $\matrixel{f}{\hat{\mathcal{T}}_2}{i}$.
First, in the transition amplitude of $\mathcal{T}_1$, it is not difficult to obtain that only $\hat{V}_{AB}$ contributes to transition amplitude,
\begin{align}
    \nonumber
    \bra{{f}}\hat{\mathcal{T}}_1\ket{{i}}=&\bra{{f}}\hat{V}_{AB}\ket{{i}}\\
    =&
    \bra{a;b'}
    \sum_{\xi\in A}
    \sum_{\zeta\in B}
    \frac{q_\xi q_{\zeta}}
    {4\pi\epsilon_0\abs{\hat{\mathbf{r}}_\xi-\hat{\mathbf{r}}_{\zeta}}}
    \ket{a';b},
    \label{Eq:T1}
\end{align}
where $\epsilon_0$ denotes the vacuum permittivity.
Equation (\ref{Eq:T1}) clearly shows that the first-order perturbation excludes the interplay of molecules and polaritons
due to $\hat{V}_{\mathrm{pol},A}$ ($\hat{V}_{\mathrm{pol},B}$) as indirect couplings between two entities. 
This is the reason that the second-order perturbation is required in our theory.
Second, the transition amplitude of $\mathcal{T}_2$ contains the following non-zero terms,
\begin{align}
    \nonumber
    \bra{{f}}\hat{\mathcal{T}}_2\ket{i}
    &= \bra{f}\hat{V}_{\mathrm{pol},A}\hat{\mathcal{G}}_0 \hat{V}_{\mathrm{pol},B}\ket{{i}}
    + \bra{f}\hat{V}_{\mathrm{pol},B}\hat{\mathcal{G}}_0
    \hat{V}_{\mathrm{pol},A}\ket{{i}}\\
    &~+\bra{f}\hat{V}_{AB}\hat{\mathcal{G}}_0\hat{V}_{AB}\ket{i}
    .
    \label{Eq:defT2}
\end{align}
It is worth noting that  $\bra{f}\hat{V}_{AB}\hat{\mathcal{G}}_0\hat{V}_{AB}\ket{i}$ can be neglect if the entities are far apart, resulting in negligible Coulomb interactions.
At the present stage, we neglect the contribution of $\bra{f}\hat{V}_{AB}\hat{\mathcal{G}}_0\hat{V}_{AB}\ket{i}$ and evaluate the second-order transition amplitude by using the spectral representation of the retarded Green's operator, 
\begin{widetext}
\begin{align}
    \nonumber
    \bra{{f}}\hat{\mathcal{T}}_2\ket{i}&\approx\bra{f}\hat{V}_{\mathrm{pol},A}\hat{\mathcal{G}}_0 \hat{V}_{\mathrm{pol},B}\ket{{i}} + \bra{{f}}\hat{V}_{\mathrm{pol},B}
    \hat{\mathcal{G}}_0\hat{V}_{\mathrm{pol},A}\ket{{i}}\\
    \nonumber
    &= \sum_{l=1}^3\int \dd[3]{\vb{s}}\int_0^\infty \dd{\omega}   \frac{\bra{a;\left\{0\right\}}\hat{V}_{\mathrm{pol},A}\ket{a';\left\{1_l(\vb{s},\omega)\right\}}
    \bra{b';\left\{1_l(\vb{s},\omega)\right\}}\hat{V}_{\mathrm{pol},B}\ket{b;\left\{0\right\}}}{E_{b}-E_{b'}-\hbar\omega+i\eta}\\
    &~+ \sum_{l=1}^3\int \dd[3]{\vb{s}}\int_0^\infty \dd{\omega}   \frac{\bra{b';\left\{0\right\}}\hat{V}_{\mathrm{pol},B}\ket{b;\left\{1_l(\vb{s},\omega)\right\}}
    \bra{a;\left\{1_l(\vb{s},\omega)\right\}}\hat{V}_{\mathrm{pol},A}\ket{a';\left\{0\right\}}}{E_{a'}-E_{a}-\hbar\omega+i\eta}.
    \label{Eq:HGH1}
\end{align}
\end{widetext}

\begin{figure}[!ht]
    \hspace{10 mm}%
    \centering
    \includegraphics[width=.33\textwidth]{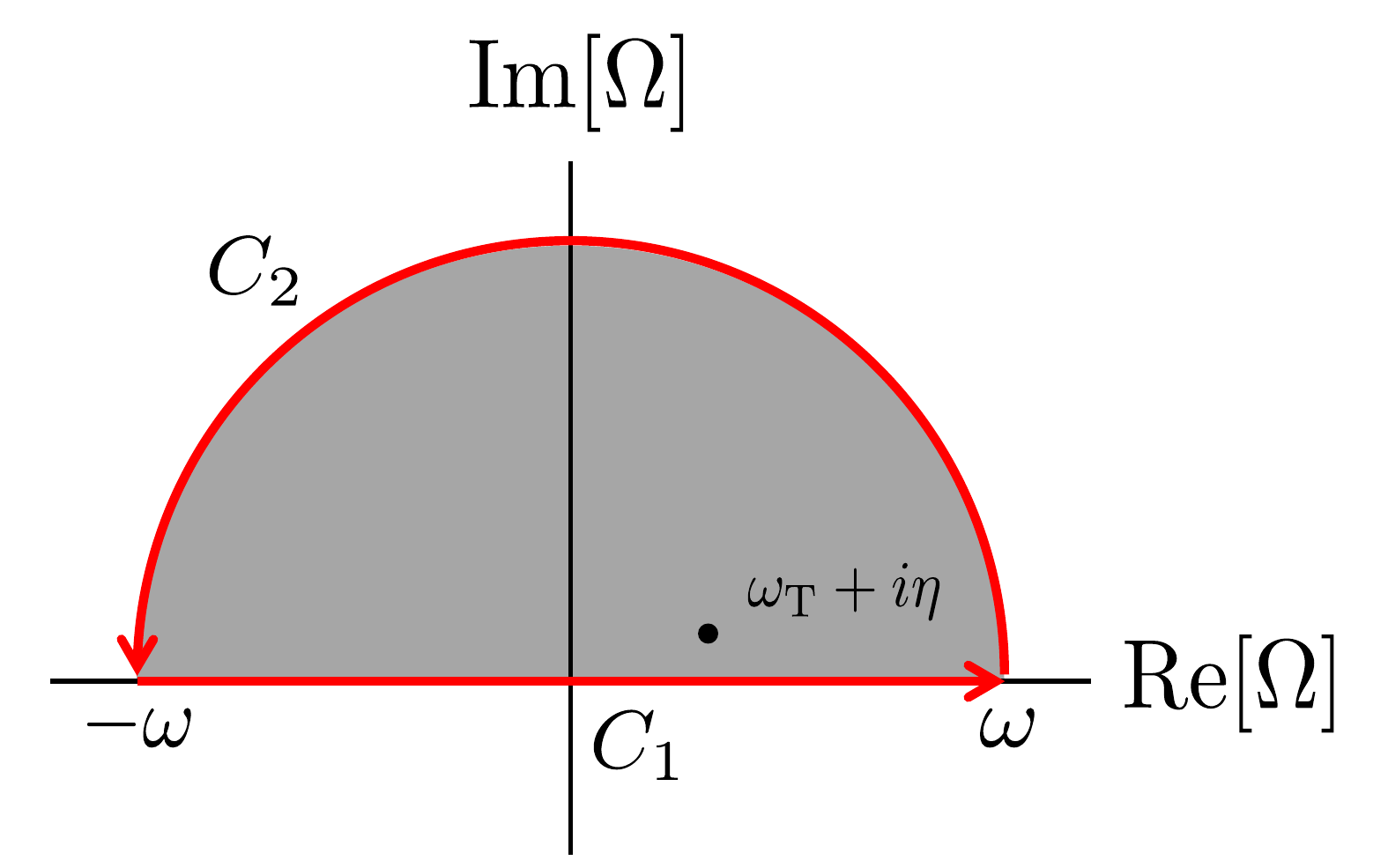}
    \caption{Illustration of the contour adopted in the $\omega$-integral. The total closed contour $C$ is equal to $C_1+C_2$. A singularity at $\Omega=\omega_\mathrm{T}+i\eta$ is located in the upper complex half plane.}
    \label{Fig:contour}%
\end{figure}

\noindent
In $\bra{f}\hat{V}_{\mathrm{pol},A}\hat{\mathcal{G}}_0 \hat{V}_{\mathrm{pol},B}\ket{{i}}$ and $\bra{{f}}\hat{V}_{\mathrm{pol},B}
\hat{\mathcal{G}}_0\hat{V}_{\mathrm{pol},A}\ket{{i}}$, we consider the following two intermediate states, $\ket{a',b';\left\{1_l(\vb{s},\omega)\right\}}$ and $\ket{a,b;\left\{1_l(\vb{s},\omega)\right\}}$, respectively. 
Recall that the single-polariton Fock state in Eq.~(\ref{Eq:HGH1}) is defined by $\ket{\left\{1_l(\vb{s},\omega)\right\}}=f_l^\dagger(\vb{s},\omega)\ket{\left\{0\right\}}$,
which is interpreted as single-polariton density with the polarization component $l$ at the frequency $\omega$ and at the position $\vb{s}$.
Furthermore, to adequately cope with $\hat{V}_{\mathrm{pol},M}$, one can define the auxiliary current density of the entity $M$,
\begin{subequations}
\begin{align}
    &{\mathcal{J}}_{mm'}(\vb{r};\omega)\equiv
    \frac{\omega}{\omega_{m'm}}
    {\vb{j}}_{mm'}^\parallel(\vb{r})+
    {\vb{j}}_{mm'}^\perp(\vb{r}),\\
    &{\mathcal{J}}_{m'm}(\vb{r};\omega)\equiv
    \frac{\omega}{\omega_{m'm}}
    {\vb{j}}_{m'm}^\parallel(\vb{r})+
    {\vb{j}}_{m'm}^\perp(\vb{r}),
\end{align}
\label{Eq:AuxJdef}%
\end{subequations}
where $\omega_{m'm}=\omega_{m'}-\omega_{m}$ and ${\vb{j}}_{mm'(m'm)}^{\parallel(\perp)}(\vb{r})$ as the longitudinal (transverse) part of the transition current density ${\vb{j}}_{mm'(m'm)}(\vb{r})=\matrixel{m(m')}{\hat{\vb{j}}_M(\vb{r})}{m'(m)}$.
For convenience, we restrict the denominator $\omega_{m'm}$ in Eq.~(\ref{Eq:AuxJdef}) to be positive and $\omega_{mm'}$ to be negative.
Therefore, according to the auxiliary transition current density, each element of $\hat{V}_{\mathrm{pol},M}$ in Eq.~(\ref{Eq:HGH1}) can be expressed as 
\\
\begin{widetext}
\begin{subequations}
    \begin{align}
        &\bra{a;\left\{0\right\}}
        \hat{V}_{\mathrm{pol},A}
        \ket{a';\left\{1_l(\vb{s},\omega)\right\}}=
        -\int\dd[3]{\vb{r}}\int_0^\infty \dd{\omega'}
        (i\omega')^{-1}
        \bra{\left\{0\right\}}\hat{\mathbf{E}}(\mathbf{r},\omega')\ket{\left\{1_l(\vb{s},\omega)\right\}}\cdot{\mathcal{J}}_{aa'}(\vb{r};-\omega'),\\
        &\bra{b';\left\{1_l(\vb{s},\omega)\right\}}
        \hat{V}_{\mathrm{pol},B}
        \ket{b;\left\{0\right\}}=
        \int\dd[3]{\vb{r}}\int_0^\infty \dd{\omega'}
        (i\omega')^{-1}
        \bra{\left\{1_l(\vb{s},\omega)\right\}}\hat{\mathbf{E}}^\dagger(\mathbf{r},\omega')\ket{\left\{0\right\}}\cdot{\mathcal{J}}_{b'b}(\vb{r};-\omega'),\\
        &\bra{b';\left\{0\right\}}
        \hat{V}_{\mathrm{pol},B}
        \ket{b;\left\{1_l(\vb{s},\omega)\right\}}=
        -\int\dd[3]{\vb{r}}\int_0^\infty \dd{\omega'}
        (i\omega')^{-1}
        \bra{\left\{0\right\}}\hat{\mathbf{E}}(\mathbf{r},\omega')\ket{\left\{1_l(\vb{s},\omega)\right\}}\cdot{\mathcal{J}}_{b'b}(\vb{r};\omega'),\\
        &\bra{a;\left\{1_l(\vb{s},\omega)\right\}}
        \hat{V}_{\mathrm{pol},A}
        \ket{a';\left\{0\right\}}=
        \int\dd[3]{\vb{r}}\int_0^\infty \dd{\omega'}
        (i\omega')^{-1}
        \bra{\left\{1_l(\vb{s},\omega)\right\}}\hat{\mathbf{E}}^\dagger(\mathbf{r},\omega')\ket{\left\{0\right\}}\cdot{\mathcal{J}}_{aa'}(\vb{r};\omega'),
\end{align}
\label{Eq:HABint}%
\end{subequations}
where the details can be found in Appendix \ref{sect:coupling}.
Note that the integral variable $\omega'$ differs from the frequency $\omega$ in the single-polariton Fock state.
Substituting Eqs.~(\ref{Eq:Edef}) and (\ref{Eq:HABint}) into Eq.~(\ref{Eq:HGH1}) with mathematical operations and using the identity to contract two dyadic Green's functions,
\begin{align}
    \mathrm{Im}~\tensor{\vb{G}}(\vb{r},\vb{r}',\omega)
    &=\int\dd[3]{\vb{s}}\frac{\omega^2\epsilon_\mathrm{I}(\vb{s},\omega)}{c^2}~
    \tensor{\vb{G}}(\vb{r},\vb{s},\omega)\cdot
    \tensor{\vb{G}}\vphantom{G}^\dagger(\vb{r}',\vb{s},\omega),
    \label{Eq:2Gidentity}
\end{align}
we finally obtain that the transition amplitude of $\mathcal{T}_2$ becomes 
\begin{align}
    \nonumber
    \bra{{f}}\hat{\mathcal{T}}_2\ket{i}
    &= 
    \frac{\hbar}{\pi\epsilon_0c^2}
    \int\dd[3]{\vb{r}}\int \dd[3]{\vb{r}'}
    \int_0^\infty \dd{\omega}\\
    &~\times\left\{
    \frac{
    {\mathcal{J}}_{aa'}(\vb{r};-\omega)
    \cdot\mathrm{Im}~\tensor{\vb{G}}(\vb{r},\vb{r}',\omega)\cdot
    {\mathcal{J}}_{b'b}(\vb{r}';-\omega)}
    {\hbar(\omega_{b}-\omega_{b'}-\omega)+i\eta}+
    \frac{
    {\mathcal{J}}_{b'b}(\vb{r};\omega)
    \cdot\mathrm{Im}~\tensor{\vb{G}}(\vb{r},\vb{r}',\omega)\cdot
    {\mathcal{J}}_{aa'}(\vb{r}';\omega)}
    {\hbar(\omega_{a'}-\omega_{a}-\omega)+i\eta}\right\}.
    \label{Eq:T2ImG1}
\end{align}
\end{widetext}
To further simplify the transition amplitude in Eq.~(\ref{Eq:T2ImG1}), we next evaluate the $\omega$-integral by contour integration in the complex domain.
We evaluate the $\omega$-integral by extending the interval $[0,\infty)$ to the whole real axis through the transformation  $\omega\rightarrow-\omega$ to the first term in Eq.~(\ref{Eq:T2ImG1}) and using the identity $\mathrm{Im}~\tensor{\vb{G}}(\vb{r},\vb{r}',-\omega)=-\mathrm{Im}~\tensor{\vb{G}}(\vb{r},\vb{r}',\omega)$~\cite{dung2002intermolecular,Buhmann2012}.
Next, by exchanging the inner product order of the dyadic Green's function (by Onsager reciprocity~\cite{Buhmann2012}), we extend the $\omega$-integral to the whole real axis,
\begin{align}
    \nonumber
    \bra{{f}}\hat{\mathcal{T}}_2\ket{i}
    &= 
    \frac{\hbar}{\pi\epsilon_0c^2}
    \int\dd[3]{\vb{r}}\int \dd[3]{\vb{r}'}
    \int_{-\infty}^\infty \dd{\omega}\\
    \label{Eq:T2temp}
    &~\times
    \frac{
    {\mathcal{J}}_{b'b}(\vb{r};\omega)
    \cdot\mathrm{Im}~\tensor{\vb{G}}(\vb{r},\vb{r}',\omega)\cdot
    {\mathcal{J}}_{aa'}(\vb{r}';\omega)}
    {\hbar(\omega_{\mathrm{T}}-\omega)+i\eta\sgn{\omega}},
\end{align}
where $\sgn{z}\equiv z/\abs{z}$ is a sign function.
Here, we assume that the transition frequency of entities $A$ and $B$ is the same, $\omega_{\mathrm{T}}=\omega_{a'}-\omega_{a}=\omega_{b'}-\omega_{b}$, because we focus on the process of ``resonance" energy transfer.
Moreover, because $\mathrm{Im}~\tensor{\vb{G}}(\vb{r},\vb{r}',\omega)$ is not holomorphic, we rewrite Eq.~(\ref{Eq:T2temp}) as
\begin{align}
    \bra{{f}}\hat{\mathcal{T}}_2\ket{i}
    &= 
    \frac{\hbar}{\pi\epsilon_0}
    \int\dd[3]{\vb{r}}\int \dd[3]{\vb{r}'}
    \int_{-\infty}^\infty \dd{\omega}I(\vb{r},\vb{r}',\omega),
\end{align}
where
\begin{align}
    \nonumber
    &I(\vb{r},\vb{r}',\omega)\\
    \nonumber
    &=\frac{1}{2ic^2}
    \Bigg\{
    \frac{
    {\mathcal{J}}_{b'b}(\vb{r};\omega)
    \cdot 
    \tensor{\vb{G}}(\vb{r},\vb{r}',\omega)
    \cdot
    {\mathcal{J}}_{aa'}(\vb{r}';\omega)}
    {\hbar(\omega_{\mathrm{T}}-\omega)+i\eta\sgn{\omega}}\\
    &\hspace{1 cm}-\frac{
    {\mathcal{J}}_{b'b}(\vb{r};\omega)
    \cdot 
    \tensor{\vb{G}}\vphantom{G}^*(\vb{r},\vb{r}',\omega)
    \cdot
    {\mathcal{J}}_{aa'}(\vb{r}';\omega)}
    {\hbar(\omega_{\mathrm{T}}-\omega)+i\eta\sgn{\omega}}
    \bigg\}.
    \label{Eq:Idef}
\end{align}
By using the fact that the dyadic Green's function $\tensor{\vb{G}}\vphantom{G}(\vb{r},\vb{r}',\omega)$ is holomorphic in the upper complex half plane~\cite{dung2002intermolecular}, we choose the path, as shown in Fig.~\ref{Fig:contour}, and evaluate the $\omega$-integral.
Hence, the $\omega$-integral becomes
\begin{align}
    \int_{-\infty}^{\infty} I(\vb{r},\vb{r}',\omega) \dd{\omega}=
    \oint_{C} I(\vb{r},\vb{r}',\Omega) \dd{\Omega}-
    \int_{C_2} I(\vb{r},\vb{r}',\Omega) \dd{\Omega}.
    \label{Eq:T2contour}%
\end{align}
According to the result derived in Appendix \ref{sect:Gamma2}, the contour integral gives the result
\begin{align}
    \label{Eq:T2Gamma2}
    \left.\bra{{f}}\hat{\mathcal{T}}_2\ket{i}\right|_{C_2}
    &= 
    \frac{1}{\epsilon_0\omega_\mathrm{T}^2}
    \int\dd[3]{\vb{r}}
    {\vb{j}}_{b'b}^\parallel(\vb{r})\cdot
    {\vb{j}}_{aa'}^\parallel(\vb{r}),\\
    &=
    \bra{a;b'}
    \hat{V}_{AB}
    \ket{a';b}.
    \label{Eq:TA_Gamma2}
\end{align}
In Eq.~(\ref{Eq:T2Gamma2}), the coupling of longitudinal transition current densities between entities $A$ and $B$ can be further reduced to the Coulomb interaction form (the derivation details can be found in Appendix~\ref{sect:longcurrent}), as a consequence of the Coulomb gauge. For the contour integral of the closed path $C=C_1+C_2$, according to the residue theorem, we can obtain
\begin{align}
    \nonumber
    \oint_{C} I(\vb{r},\vb{r}',\Omega) \dd{\Omega}
    &= 2\pi i~\mathrm{Res}[I(\vb{r},\vb{r}',\Omega),\omega_\mathrm{T}+i\eta/\hbar]\\
    \nonumber
    &=\frac{\pi}{\hbar c^2}
    {\vb{j}}_{b'b}(\vb{r})
    \cdot \tensor{\vb{G}}(\vb{r},\vb{r}',\omega_{\mathrm{T}})\cdot
    {\vb{j}}_{aa'}(\vb{r}'),
\end{align}
which indicates that 
\begin{align}
    \nonumber
    \left.\bra{{f}}\hat{\mathcal{T}}_2\ket{i}\right|_{C}&= 
    \frac{1}{\epsilon_0c^2}
    \int\dd[3]{\vb{r}} \int\dd[3]{\vb{r}'}\\
    &~\times
    {\vb{j}}_{b'b}(\vb{r})
    \cdot \tensor{\vb{G}}(\vb{r},\vb{r}',\omega_{\mathrm{T}})\cdot
    {\vb{j}}_{aa'}(\vb{r}').
    \label{Eq:T2Gamma}
\end{align}
Recall that the auxiliary current density defined in Eq.~(\ref{Eq:AuxJdef}) gives 
${\mathcal{J}}_{b'b}(\vb{r};\omega_{\mathrm{T}})={\vb{j}}_{b'b}(\vb{r})$ and ${\mathcal{J}}_{aa'}(\vb{r}';\omega_{\mathrm{T}})={\vb{j}}_{aa'}(\vb{r}')$.
Finally, according to Eqs.~(\ref{Eq:T1}), (\ref{Eq:T2contour}), (\ref{Eq:T2Gamma2}), and (\ref{Eq:T2Gamma}), we obtain the total transition amplitude,
\begin{align}
    \nonumber
    \bra{{f}}\hat{\mathcal{T}}\ket{{i}} 
    &=\frac{1}{\epsilon_0c^2}
    \int\dd[3]{\vb{r}} \int\dd[3]{\vb{r}'}\\
    &~\times
    {\vb{j}}_{b'b}(\vb{r})
    \cdot \tensor{\vb{G}}(\vb{r},\vb{r}',\omega_{\mathrm{T}})\cdot
    {\vb{j}}_{aa'}(\vb{r}').
    \label{Eq:TAFinal}
\end{align}
Note that the transition amplitudes in Eqs.~(\ref{Eq:T1}) and (\ref{Eq:T2Gamma2}) mutually cancel out.
Finally, we obtain the RET rate in terms of transition current density in the following expression, 
\begin{widetext}
\begin{align}
    \Gamma &= \frac{2\pi}{\hbar^2}\int_0^\infty \dd{\omega}
    \sum_{(a,b')}\sum_{(a',b)} P_{a'}P_{b}
    \abs{
    \frac{1}{\epsilon_0c^2}
    \int\dd[3]{\vb{r}} \int\dd[3]{\vb{r}'}
    {\vb{j}}_{b'b}(\vb{r})
    \cdot \tensor{\vb{G}}(\vb{r},\vb{r}',\omega)\cdot
    {\vb{j}}_{aa'}(\vb{r}')
    }^2
    \delta(\omega_{a'}-\omega_a-\omega)\delta(\omega_{b'}-\omega_b-\omega),
    \label{Eq:RETfin}
\end{align}
\end{widetext}
where $(a',b)$ [$(a,b')$] denotes the grouped indices of initial (final) state, and  $P_{a'}$($P_b$) denotes the initial-state probability distribution of the entity $A$($B$).
Recall that the transition frequency of $A$ and $B$ are the same, $\omega_{a'}-\omega_{a}=\omega_{b'}-\omega_{b}$.
In the current stage, we have derived an explicit form of the RET rates in terms of transition current density, as shown in Eq.~(\ref{Eq:RETfin}), but this equation cannot be used to directly evaluate RET rates in material systems via \textit{ab initio} methods. To solve this issue, we adopt a further approximation in the next section.

\subsection{Transition Current Density and Molecule}
To evaluate the transition current density via \textit{ab initio} methods, it is necessary to separate the electronic and nuclear degrees of freedom first (approach to recapturing the vibronic effect can be found in Ref.~\cite{Nafie1997,Freedman1998,Freedman1997}).
In the same spirit of the Condon approximation~\cite{Condon1928,Nitzan2006}, we approximate the transition current density to
\begin{align}
    \nonumber
    \vb{j}_{mm'}(\vb{r})&=\matrixel{m}{\hat{\vb{j}}_M(\vb{r})}{m'}\\
    &\approx
    \matrixel*{\phi_{M,\gamma}}{\hat{\vb{j}}_M(\vb{r})}{\phi_{M,\gamma'}}_{\{\vb{R}\}}
    \braket{\chi_{M,\nu}}{\chi_{M,\nu'}},
    \label{Eq:CondonlikeApprox}
\end{align}
where $\ket*{\phi_{M,\gamma}}$ denotes the $\gamma$-th electronic state, and $\ket{\chi_{M,\nu}}$ is the $\nu$-th nuclear state associated with the $\gamma$-th electronic state.
Here, the subscript $\{\vb{R}\}$ represents the electronic element is evaluated under a specific nuclear coordinates $\{\vb{R}\}$.
The Condon-like approximation allows us to separate the electronic and nuclear degrees of freedoms. As a result, we can focus only on the electronic transition current density, and the contribution of the nuclear part can be attributed to the nuclear wavefunction overlap $\braket{\chi_{M,\nu}}{\chi_{M,\nu'}}$.
As a consequence, the quantum number $m$ is now assigned by the two indices, $m\rightarrow(\gamma,\nu)$.
In addition, because most \textit{ab initio} calculations are performed in the coordinate space, the projection of the states to the position-spin coordinates is required.
After taking the projection and considering the antisymmetric property of electrons, we finally get the electronic transition current density (detail derivations can be found in Appendix \ref{sect:TCD})
\begin{align}
    \nonumber
    \vb{j}_{\gamma\gamma'}^M(\vb{r})
    &\equiv\matrixel*{\phi_{M,\gamma}}{\hat{\vb{j}}_M(\vb{r})}{\phi_{M,\gamma'}}_{\{\vb{R}\}}\\
    \nonumber
    &=
    \frac{-i\hbar e N_\mathrm{el}}{2m_\mathrm{el}}
    \bigg[
    \tilde{\phi}_{M,\gamma}^*(\vb{r};\{\vb{R}\})\nabla\tilde{\phi}_{M,\gamma'}(\vb{r};\{\vb{R}\})\\
    &\hspace{1.3 cm}-
    \tilde{\phi}_{M,\gamma'}(\vb{r};\{\vb{R}\})\nabla\tilde{\phi}_{M,\gamma}^*(\vb{r};\{\vb{R}\})
    \bigg],
    \label{Eq:j_ewfn}
\end{align}
where $N_\mathrm{el}$ is the total number of electrons in molecule $M$, $e$ is the elementary charge, and $m_\mathrm{el}$ is the electron mass.
Remember that the gradient operator only operates on $\vb{r}$, not $\{\vb{R}\}$.
Moreover, $\tilde{\phi}_{M,\gamma}(\vb{r},\{\vb{R}\})$ is the single-electron wavefunction of the electronic state $\gamma$, which is defined by
\begin{align}
    \nonumber
    &\tilde{\phi}_{M,\gamma}^*(\vb{x}_1;\{\vb{R}\})\tilde{\phi}_{M,\gamma'}(\vb{x}_1;\{\vb{R}\})\\
    &\equiv
    \int \mathcal{D}\{\vb{x}_{\mu\neq1}\}~\phi_{M,\gamma}^*(\{\vb{x}_\mu\};\{\vb{R}\})\phi_{M,\gamma'}(\{\vb{x}_\mu\};\{\vb{R}\}).
    \label{Eq:eff1ewfn}
\end{align}
In Eq.~(\ref{Eq:eff1ewfn}), $\{\vb{x}_\mu\}$ denotes the set of position-spin coordinates, and $\phi_{M,\gamma}(\{\vb{x}_\mu\};\{\vb{R}\})$ denotes the multi-electron wavefunction of the electronic state $\gamma$, which is parameterized by a specific nuclear coordinates $\{\vb{R}\}$. The integration symbol represents a series of integrals except for $\vb{x}_1$,
\begin{align}
    \nonumber
    \int \mathcal{D}\{\vb{x}_{\mu\neq1}\}\equiv
    \int \dd{\vb{x}_2} \dots \int \dd{\vb{x}_{N_\mathrm{el}}}.
\end{align}

\begin{figure*}[!t]
    \centering
    \includegraphics[width=1\textwidth]{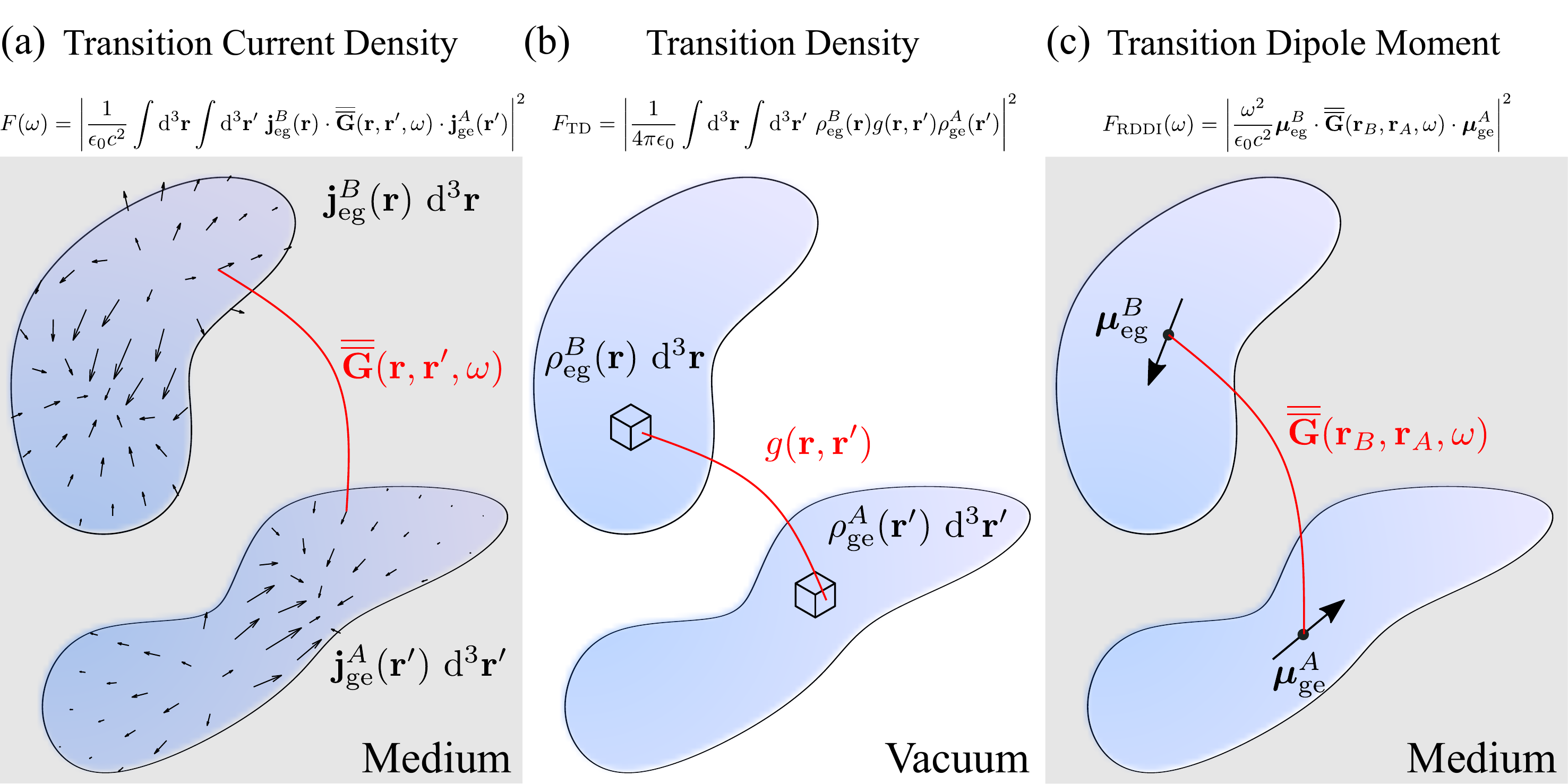}
    \caption{Schematic diagram of the electromagnetic coupling factors in three different theoretical approaches: (a) transition current density, (b) transition density, and (c) transition dipole moment.}
    \label{Fig:Illust}
\end{figure*}

Specially, in molecular systems, it is common to approximate the single-electron wavefunctions as molecular orbitals (MO), which can be obtained from \textit{ab initio} calculations.
If the transition between the electronic ground state and the first excited state plays the most important transition,
one can assume that the single-electron wavefunction can be properly described by
\begin{subequations}
\begin{align}
    \tilde{\phi}_{M,\mathrm{e}}(\vb{r};\{\vb{R}\})&\approx
    \phi_{M,\mathrm{LUMO}}(\vb{r};\{\vb{R}\}),
    &\gamma' \equiv \mathrm{e},\\
    \tilde{\phi}_{M,\mathrm{g}}(\vb{r};\{\vb{R}\})&\approx
    \phi_{M,\mathrm{HOMO}}(\vb{r};\{\vb{R}\}),
    &\gamma \equiv \mathrm{g},
\end{align}
\label{Eq:MOapprox}%
\end{subequations}
where LUMO and HOMO are the abbreviation of the lowest unoccupied MO and the highest occupied MO, respectively.
In other words, the transition current density for the molecule $M$ is determined by its HOMO, LUMO, and the nuclear wavefunction overlap.
Under the approximation in Eq.~(\ref{Eq:MOapprox}), the transition current density can be expressed as
\begin{align}
    \vb{j}_{mm'}(\vb{r})&=
    \braket{\chi_{M,\nu}}{\chi_{M,\nu'}}~
    \vb{j}_\mathrm{ge}^M(\vb{r}).
\end{align}
According to Eqs.~(\ref{Eq:j_ewfn}) and~(\ref{Eq:MOapprox}), one can obtain $\vb{j}_\mathrm{ge}^M(\vb{r})$ as follows, 
\begin{align}
    \nonumber
    \vb{j}_\mathrm{ge}^M(\vb{r}) &= 
    \frac{-i\hbar e N_\mathrm{el}}{2m_\mathrm{el}}\\
    \nonumber
    &~\times
    \bigg[
    \phi_{M,\mathrm{HOMO}}^*(\vb{r};\{\vb{R}\})\nabla\phi_{M,\mathrm{LUMO}}(\vb{r};\{\vb{R}\})\\
    &\hspace{.31 cm}-
    \phi_{M,\mathrm{LUMO}}(\vb{r};\{\vb{R}\})\nabla\phi_{M,\mathrm{HOMO}}^*(\vb{r};\{\vb{R}\})
    \bigg].
\end{align}
Eventually, the RET rates in Eq.~(\ref{Eq:RETfin}) can be expressed and interpreted as a generalized spectral overlap between two molecules and electromagnetic coupling factor $F(\omega)$,
\begin{align}
    \Gamma &= \frac{2\pi}{\hbar^2}\int_0^\infty \dd{\omega}
    \mathcal{L}_B^{\mathrm{abs}}(\omega)
    F(\omega)
    \mathcal{L}_A^{\mathrm{em}}(\omega),
    \label{Eq:RETCondonfin}
\end{align}
with
\begin{align}
    \nonumber
    \mathcal{L}_A^{\mathrm{em}}(\omega)&\equiv
    \sum_{a,a'}
    P_{a'}
    \abs{\braket{\chi_{A,\alpha}}{\chi_{A,\alpha'}}}^2
    \delta(\omega_{a'}-\omega_a-\omega),\\
    \nonumber
    \mathcal{L}_B^{\mathrm{abs}}(\omega)&\equiv
    \sum_{b,b'} 
    P_{b}
    \abs{\braket{\chi_{B,\beta'}}{\chi_{B,\beta}}}^2
    \delta(\omega_{b'}-\omega_b-\omega),
\end{align}
and
\begin{align}
    \nonumber
    F(\omega)
    &=
    \abs{\frac{1}{\epsilon_0c^2}
    \int\dd[3]{\vb{r}} \int\dd[3]{\vb{r}'}
    {\vb{j}}_\mathrm{eg}^B(\vb{r})
    \cdot \tensor{\vb{G}}(\vb{r},\vb{r}',\omega)\cdot
    {\vb{j}}_\mathrm{ge}^A(\vb{r}')}^2.
\end{align}
The schematic illustration of $F(\omega)$ is shown in Fig.~\ref{Fig:Illust}a.
Note that $\mathcal{L}_B^{\mathrm{abs}}(\omega)$ and $\mathcal{L}_A^{\mathrm{em}}(\omega)$ are related to the absorption spectrum of $B$ and the emission spectrum of $A$, respectively.

\section{Applicability to Former Theories}
\label{sect:app}
In Sec.~\ref{sect:theory}, we have shown how to derive the RET rate based on the transition current density approach in the framework of macroscopic QED.
Furthermore, to demonstrate the generality of our theory, in this section, we will prove that  Eq.~(\ref{Eq:RETCondonfin}) can recover to the main results in previous studies: the TDC method~\cite{Krueger1998} and the plasmon-coupled resonance energy transfer~\cite{Hsu2017}.


\subsection{Transition Density Cube Method}
When two entities are close, Krueger \textit{et al.} have developed the TDC method \cite{Krueger1998} in the electrostatic limit and demonstrated how to calculate the Coulomb coupling between the pigments of the bacterial light-harvesting complex.
To recover the main result given in the TDC method, we consider the free-space dyadic Green's function $\tensor{\vb{G}}_0(\vb{r},\vb{r}',\omega)$ and adopt the electrostatic approximation,
\begin{align}
    \tensor{\vb{G}}(\vb{r},\vb{r}',\omega)\rightarrow
    \lim_{k_0\rightarrow 0} \tensor{\vb{G}}_0(\vb{r},\vb{r}',\omega)=
    \nabla\nabla\frac{1}{4\pi k_0^2R},
\end{align}
where $k_0=\omega/c$, $R=\abs{\vb{r}-\vb{r}'}$, and $\nabla\nabla$ is a dyadic operator.
Thus, under the electrostatic limit, the electromagnetic coupling factor $F(\omega)$  becomes $F_\mathrm{TD}$ (TD is the acronym for the transition density) as follows, 
\begin{align}
    \nonumber
    F_\mathrm{TD}&\equiv\lim_{k_0\rightarrow0}F(\omega)\\
    \nonumber
    &=\abs{\frac{1}{\epsilon_0 c^2}
    \int\dd[3]{\vb{r}} \int\dd[3]{\vb{r}'}
    \nabla\cdot{\vb{j}}_\mathrm{eg}^B(\vb{r})
    \frac{1}{4\pi k_0^2R}
    \nabla'\cdot{\vb{j}}_\mathrm{ge}^A(\vb{r}')}^2\\
    &= \abs{\frac{1}{4\pi\epsilon_0}
    \int\dd[3]{\vb{r}} \int\dd[3]{\vb{r}'}
    \rho_\mathrm{eg}^B(\vb{r})g(\vb{r},\vb{r}')\rho_\mathrm{ge}^A(\vb{r}')}^2,
    \label{Eq:TDCcoupling}
\end{align}
where $\rho_\mathrm{ge}^A(\vb{r}')$ and $\rho_\mathrm{eg}^B(\vb{r})$ are transition densities of the entity A and the entity B, respectively, and $g(\vb{r},\vb{r}')=1/\abs{\vb{r}-\vb{r}'}$ is the scalar Green's function of the Poisson equation.
Note that the derivation of Eq.~(\ref{Eq:TDCcoupling}) requires the continuity equation.
Obviously, $F_\mathrm{TD}$ is exactly the continuous form of the electronic coupling corresponding to Eq.~(5) in Ref.~\cite{Krueger1998}.
Incidentally, to numerically perform the integral in Eq.~(\ref{Eq:TDCcoupling}), Krueger \textit{et al.} discretized the space into sufficiently small cubes shown in Fig.~\ref{Fig:Illust}b, hence the name transition density cube (TDC).
Furthermore, because $F_\mathrm{TD}$ is independent of frequency, it can be taken out from the generalized spectral overlap, and the RET rate becomes
\begin{align}
    \Gamma = \frac{2\pi}{\hbar^2}F_\mathrm{TD}J,
    \qquad J = \int_0^\infty\dd{\omega}
    \mathcal{L}_B^{\mathrm{abs}}(\omega)
    \mathcal{L}_A^{\mathrm{em}}(\omega),
    \label{Eq:TDCrate}
\end{align}
where $F_\mathrm{TD}$ corresponds to $\abs{V}^2$ in Eq.~(8) in Ref.~\cite{Krueger1998}.
Note that the prefactor of Eq.~(8) in Ref.~\cite{Krueger1998} is slightly different from that in Eq.~(\ref{Eq:TDCrate}) because of their spectral overlap integrals in different units ($J$ in Ref.~\cite{Krueger1998} expressed in $\bar{\nu} = \omega/2\pi c$).
In addition, we would like to point out that $J$ defined in Eq.~(\ref{Eq:TDCrate}) is associated with the well-known spectral overlap in F\"orster's theory.
In brief, the TDC method is the electrostatic limit of our theory in free space (i.e., neglect
of the retardation effect in homogeneous, non-dispersive, and non-absorbing media).

\subsection{Plasmon-Coupled Resonance Energy Transfer}


Equation~(\ref{Eq:RETCondonfin}) is general for us to describe RET in linear, dispersive, and absorbing media. In other words, the effect of plasmon polaritons can be included in our theory, indicating that the main result of plasmon-coupled RET should be able to be recovered.
The influence of plasmon polaritons on the RET rates under the EDA have been discussed in various electromagnetic environments~\cite{Ding2017,Ding2018,Wu2018,Lee2020,Wei2021}.
To obtain the RET rate [Eq.~(2)] in Ref.~\cite{Hsu2017}, we start from the definition of the current density and apply the relation $\hat{\vb{p}}_\xi=\comm*{\hat{\vb{r}}_\xi}{\hat{H}_M}$,
\begin{align}
    \nonumber
    {\hat{\vb{j}}}_M(\vb{r})
    &=
    \sum_{\xi\in M}\frac{q_\xi}{2i\hbar}\bigg\{
    \delta(\vb{r}-\hat{\vb{r}}_\xi)\comm{\hat{\vb{r}}_\xi}{\hat{H}_M}+
    \hc
    \bigg\}\\
    \nonumber
    &\approx
    \sum_{\xi\in M}\frac{q_\xi}{2i\hbar}\bigg\{
    \delta(\vb{r}-\vb{r}_M)\comm{\hat{\vb{r}}_\xi-\vb{r}_M}{\hat{H}_M}+
    \hc
    \bigg\}\\
    &=
    \frac{1}{i\hbar}\comm{\hat{\boldsymbol{\mu}}_M}{\hat{H}_M}
    \delta(\vb{r}-\vb{r}_M).
    \label{Eq:jpdapprox}
\end{align}
Here, we introduce the center of mass $\vb{r}_M$ for the entity $M$ and make the point-dipole approximation to get the dipole operator $\hat{\boldsymbol{\mu}}_M$.
Through calculating $\bra{m(m')}{\hat{\vb{j}}}_M(\vb{r})\ket{m'(m)}=\vb{j}_{mm'(m'm)}(\vb{r})$ in Eq.~(\ref{Eq:jpdapprox}), where $\ket{m(m')}$ is the $m$-th ($m'$-th) energy eigenstate of an entity $M$, we can obtain 
\begin{align}
    \vb{j}_{mm'(m'm)}(\vb{r})=
    \mp i\omega_\mathrm{T}\boldsymbol{\mu}_{mm'(m'm)}
    \delta(\vb{r}-\vb{r}_M).
    \label{Eq:dipoleelement}
\end{align}
Next, we take the Condon approximation to the transition dipole $\boldsymbol{\mu}_{mm'(m'm)}$, it is straightforward to obtain
\begin{align}
    \boldsymbol{\mu}_{mm'(m'm)} \approx
    \boldsymbol{\mu}_{\mathrm{ge(eg)}}^M
    \braket{\chi_{M,\nu(\nu')}}{\chi_{M,\nu'(\nu)}},
    \label{Eq:CondonApprox}
\end{align}
where $\boldsymbol{\mu}_{\mathrm{ge(eg)}}^M$ is the electronic transition dipole.
Substituting Eqs.~(\ref{Eq:dipoleelement}) and (\ref{Eq:CondonApprox}) into Eq.~(\ref{Eq:RETfin}), we obtain that
\begin{align}
    \label{Eq:RETRDDI}
    \Gamma 
    &= \frac{2\pi}{\hbar^2}\int_0^\infty \dd{\omega}
    \mathcal{L}_B^{\mathrm{abs}}(\omega)
    F_\mathrm{RDDI}(\omega)
    \mathcal{L}_A^{\mathrm{em}}(\omega),
\end{align}%
where
\begin{align}
    F_\mathrm{RDDI}(\omega)=
    \abs{
    \frac{\omega^2}{\epsilon_0c^2}
    \boldsymbol{\mu}_\mathrm{eg}^B
    \cdot \tensor{\vb{G}}(\vb{r}_B,\vb{r}_A,\omega)\cdot
    \boldsymbol{\mu}_\mathrm{ge}^A
    }^2
    \label{Eq:rddi}
\end{align}%
is exactly the form of resonant dipole-dipole interaction (RDDI) in Ref.~\cite{Hsu2017}.
Here, we would like to mention that the definition of $F_\mathrm{RDDI}(\omega)$ is slightly different from that of coupling factor in Ref.~\cite{Hsu2017} because the magnitudes of the transition dipoles of molecules $A$ and $B$ in Ref.~\cite{Hsu2017} have been incorporated into the emission and absorption lineshape functions, i.e.,
\begin{subequations}
\begin{align}
    W_A^\mathrm{em}(\omega)
    &= \frac{2\pi}{\hbar^2}
    \abs{\boldsymbol{\mu}_\mathrm{ge}^A}^2
    \mathcal{L}_A^{\mathrm{em}}(\omega),\\
    W_B^\mathrm{abs}(\omega)
    &=\frac{2\pi}{\hbar^2}
    \abs{\boldsymbol{\mu}_\mathrm{eg}^B}^2
    \mathcal{L}_B^{\mathrm{abs}}(\omega).
\end{align}
\label{Eq:lineshape}%
\end{subequations}
According to Eq.~(\ref{Eq:lineshape}), it is obvious that Eq.~(\ref{Eq:RETCondonfin}) covers the main result in Ref.~\cite{Hsu2017}.
Also, it is worth mentioning that $F_\mathrm{RDDI}(\omega)$ in Eq.~(\ref{Eq:rddi}) is consistent with the transition tensor deduced from the framework of microscopic (or molecular) QED \cite{Daniels2003}, which requires the entities in the homogeneous dilute media or in vacuum. In such case, the electromagnetic coupling factor becomes
\begin{align}
    F_\mathrm{RDDI}^\mathrm{dil}(\omega)=
    \abs{
    \frac{\omega^2}{\epsilon_0c^2}
    \boldsymbol{\mu}_\mathrm{eg}^B
    \cdot\tensor{\vb{G}}\vphantom{a}^\mathrm{dil}(\vb{r}_B,\vb{r}_A,\omega)\cdot
    \boldsymbol{\mu}_\mathrm{ge}^A
    }^2,
\end{align}
with 
\begin{align}
    \nonumber
    \tensor{\vb{G}}\vphantom{a}^\mathrm{dil}(\vb{r}_B,\vb{r}_A,\omega)
    =\frac{e^{ikR}}{4\pi R}
    \bigg\{&\Big[3\hat{\vb{e}}_R\otimes\hat{\vb{e}}_R-\tensor{\vb{I}}\Big]
    \Big[\frac{1}{(kR)^2}-\frac{i}{kR}\Big]\\
    +&\Big[\tensor{\vb{I}}-\hat{\vb{e}}_R\otimes\hat{\vb{e}}_R\Big]
    \bigg\}.
\end{align}
Note that $k=n\omega/c$, $\vb{r}_B-\vb{r}_A = R~\hat{\vb{e}}_R$, and $n$ is the refractive index for dilute media.
Moreover, if we further impose the electrostatic limit on Eq.~(\ref{Eq:RETRDDI}), we can obtain the famous F\"orster theory, as discussed in Ref.~\cite{Hsu2017}.
To emphasize the difference among these RET theories, we provide a schematic diagram, shown in Fig.~\ref{Fig:Illust}, that depicts the key concepts of the three electromagnetic coupling factors in RET theories.

\begin{table*}[!t]
\caption{Summary of RET theories with and without approximations.}
\label{table:1}
\begin{ruledtabular}
\begin{tabular}{lcll}
Method & Approximation & Rate & Coupling Form\\
\hline
\footnote{Transition current density}TCD & - & 
    $\displaystyle
    \Gamma = \frac{2\pi}{\hbar^2}\int_0^\infty \dd{\omega}
    \mathcal{L}_B^{\mathrm{abs}}(\omega)
    F(\omega)
    \mathcal{L}_A^{\mathrm{em}}(\omega)$
&
    $F(\omega)=\displaystyle
    \abs{\frac{1}{\epsilon_0c^2}
    \int\dd[3]{\vb{r}} \int\dd[3]{\vb{r}'}
    {\vb{j}}_\mathrm{eg}^B(\vb{r})
    \cdot \tensor{\vb{G}}(\vb{r},\vb{r}',\omega)\cdot
    {\vb{j}}_\mathrm{ge}^A(\vb{r}')}^2$\\
\footnote{Transition density}TD & Electrostatic & 
    $\displaystyle\Gamma = \frac{2\pi}{\hbar^2}
    F_\mathrm{TD}
    \int_0^\infty\dd{\omega}
    \mathcal{L}_B^{\mathrm{abs}}(\omega)
    \mathcal{L}_A^{\mathrm{em}}(\omega)$
&
    $\displaystyle
    F_\mathrm{TD}=\abs{\frac{1}{4\pi\epsilon_0}
    \int\dd[3]{\vb{r}} \int\dd[3]{\vb{r}'}
    \rho_\mathrm{eg}^B(\vb{r})
    g(\vb{r},\vb{r}')
    \rho_\mathrm{ge}^A(\vb{r}')}^2$ \\
\footnote{Transition dipole moment (macroscopic QED)}TDM-1 & Point dipole & 
    $\displaystyle\Gamma 
    = \frac{2\pi}{\hbar^2}\int_0^\infty \dd{\omega}
    \mathcal{L}_B^{\mathrm{abs}}(\omega)
    F_\mathrm{RDDI}(\omega)
    \mathcal{L}_A^{\mathrm{em}}(\omega)$
&
    $\displaystyle
    F_\mathrm{RDDI}(\omega)=
    \abs{
    \frac{\omega^2}{\epsilon_0c^2}
    \boldsymbol{\mu}_\mathrm{eg}^B
    \cdot \tensor{\vb{G}}(\vb{r}_B,\vb{r}_A,\omega)\cdot
    \boldsymbol{\mu}_\mathrm{ge}^A
    }^2$\\
\footnote{Transition dipole moment (molecular QED)}TDM-2 & 
\makecell{Point dipole\\ (dilute media)} & 
    $\displaystyle\Gamma 
    = \frac{2\pi}{\hbar^2}\int_0^\infty \dd{\omega}
    \mathcal{L}_B^{\mathrm{abs}}(\omega)
    F_\mathrm{RDDI}^\mathrm{dil}(\omega)
    \mathcal{L}_A^{\mathrm{em}}(\omega)$
&
    $\displaystyle
    F_\mathrm{RDDI}^\mathrm{dil}(\omega)=
    \abs{
    \frac{\omega^2}{\epsilon_0c^2}
    \boldsymbol{\mu}_\mathrm{eg}^B
    \cdot 
    \tensor{\vb{G}}\vphantom{a}^\mathrm{dil}(\vb{r}_B,\vb{r}_A,\omega)\cdot
    \boldsymbol{\mu}_\mathrm{ge}^A
    }^2$\\
\footnote{F\"orster resonance energy transfer}FRET & \makecell{Electrostatic\\\& point dipole} & 
    $\displaystyle\Gamma = \frac{2\pi}{\hbar^2}
    F_\mathrm{FRET}
    \int_0^\infty\dd{\omega}
    \mathcal{L}_B^{\mathrm{abs}}(\omega)
    \mathcal{L}_A^{\mathrm{em}}(\omega)$
&
    $\displaystyle F_\mathrm{FRET}=
    \frac{\footnote{$\kappa$ is the orientation factor.}\kappa^2}{4\pi\epsilon_0R^6}
    \abs{\boldsymbol{\mu}_\mathrm{eg}^B}^2
    \abs{\boldsymbol{\mu}_\mathrm{ge}^A}^2
    $
\end{tabular}
\end{ruledtabular}
\end{table*}

\section{Conclusion}
Resonant dipole-dipole interaction has been widely used to describe light-matter interaction in physics, chemistry, and materials science; however, RDDI is only a first-step approximation because it cannot fully contain the structural information of entities (e.g., atoms, molecules, quantum dots, 2D materials) when describing light-matter interactions.
To include a spatial-dependent light-matter interaction, we developed the generalized RET theory based on the concept of transition current density in the presence of linear, dispersive, and absorbing media within the framework of macroscopic QED.
By expanding the Born series up to the second order in the time-dependent perturbation theory, we successfully derived the RET rate in the generalized-coupling expression of the transition dipole moment.
Furthermore, by applying the Condon-like approximation to the transition current density, we separated the electronic and nuclear degrees of freedom and showed that the transition current density can be described by the HOMO and LUMO, which can be obtained from \textit{ab initio} calculations.
Moreover, we expressed the RET rates in terms of the generalized spectral overlap, as shown in Eq.~(\ref{Eq:RETCondonfin}).
Finally, to demonstrate the validity and generality of Eq.~(\ref{Eq:RETCondonfin}), we proved that the present theory can be reduced to the main results in the previous studies, including TDC and plasmon-coupled RET. The comparison of several representative RET theories are summarized in Table \ref{table:1}.
In short, in the framework of macroscopic QED, the current approach provides one key step beyond the traditional RET theory based on RDDI because Eq.~(\ref{Eq:RETCondonfin}) not only serves as a generalized version (i.e., containing retardation effect) of the TDC method, but also includes the influence of photonic environments, e.g., polaritons.

The generalized RET theory beyond RDDI has been presented in this work. However, this study is just the beginning, and several issues are worth further exploration.
First, in the present theory, we do not consider the mechanism of Dexter energy transfer~\cite{Dexter1953,Skourtis2016,Murphy2004,Olaya-Castro2011} (i.e., electron exchange between two entities).
This mechanism becomes important when the wavefunction overlap of two entities cannot be negligible.
Second, the quantum dynamics of RET cannot be described in the present theory due to the limitation of Fermi's golden rule.
How to generalize the theory to include quantum dynamics is an intriguing but challenging issue.
In the end, we leave the numerical demonstration to the future study and hope that the present theory will inspire further investigation into the basic theory of energy transfer and its applications.

\begin{acknowledgments}
Hsu thanks Academia Sinica (AS-CDA-111-M02) and the Ministry of Science and Technology of Taiwan (110-2113-M-001-053 and 111-2113-M-001-027-MY4) for the financial support.
\end{acknowledgments}


\appendix

\section{Continuous Form of $\hat{\vb{A}}(\hat{\vb{r}}_\xi)\cdot\hat{\vb{p}}_\xi$ in Eq.~(\ref{Eq:Hintdef})}
\label{sect:deltap}
In this appendix, we prove that the order of momentum operator and Dirac delta function do not affect the result in the Coulomb gauge, i.e., 
\begin{align}
    \nonumber
    &\sum_{\xi\in M}\frac{q_\xi}{2m_\xi}\int \dd[3]{\vb{r}}~
    \hat{\mathbf{A}}(\mathbf{r})\cdot
    \hat{\vb{p}}_\xi\delta(\vb{r}-\hat{\vb{r}}_\xi)\\
    =&
    \sum_{\xi\in M}\frac{q_\xi}{2m_\xi}\int \dd[3]{\vb{r}}~
    \hat{\mathbf{A}}(\mathbf{r})\cdot
    \delta(\vb{r}-\hat{\vb{r}}_\xi)\hat{\vb{p}}_\xi.
\end{align}
We start from the Fourier expression of Dirac delta functions,
\begin{align}
    \delta(\vb{r}-\hat{\vb{r}}_\xi) &= \frac{1}{(2\pi\hbar)^3}\int\dd[3]{\vb{p}}
    e^{i\vb{p}\cdot(\vb{r}-\hat{\vb{r}}_\xi)/\hbar}.
    \label{Eq:deltardef}
\end{align}
According to Eq.~(\ref{Eq:deltardef}), the product of $\hat{\vb{p}}_\xi$ and $\delta(\vb{r}-\hat{\vb{r}}_\xi)$ becomes
\begin{align}
    \nonumber
    &\hat{\vb{p}}_\xi\delta(\vb{r}-\hat{\vb{r}}_\xi)\\
    &=
    \frac{1}{(2\pi\hbar)^3}
    \int\dd[3]{\vb{p}}~
    \hat{\vb{p}}_\xi\sum_{n=0}^\infty
    \frac{[i\vb{p}\cdot(\vb{r}-\hat{\vb{r}}_\xi)/\hbar]^n}{n!},
    \label{Eq:Proof1}
\end{align}
where we take the Maclaurin series of the exponential function.
Next, we exchange the order of operators $\hat{\vb{p}}_\xi$ and $\hat{\vb{r}}_\xi$. For the term of the $n$-th power, we obtain the recursive relation,
\begin{align}
    \nonumber
    &\hat{\vb{p}}_\xi[i\vb{p}\cdot(\vb{r}-\hat{\vb{r}}_\xi)/\hbar]^n\\
    &=
    \Big\{[i\vb{p}\cdot(\vb{r}-\hat{\vb{r}}_\xi)/\hbar]\hat{\vb{p}}_\xi-\vb{p}
    \Big\}[i\vb{p}\cdot(\vb{r}-\hat{\vb{r}}_\xi)/\hbar]^{n-1},
    \label{Eq:Proof2}
\end{align}
with the canonical commutation relation $\comm{\hat{r}_{\xi,j}}{\hat{p}_{\xi,k}}=i\hbar\delta_{jk}$.
Using Eq.~(\ref{Eq:Proof2}), we can show that
\begin{align}
    \nonumber
    &\hat{\vb{p}}_\xi[i\vb{p}\cdot(\vb{r}-\hat{\vb{r}}_\xi)/\hbar]^n\\
    &=
    [i\vb{p}\cdot(\vb{r}-\hat{\vb{r}}_\xi)/\hbar]^n\hat{\vb{p}}_\xi
    -n\vb{p}[i\vb{p}\cdot(\vb{r}-\hat{\vb{r}}_\xi)/\hbar]^{n-1}.
\end{align}
Hence, Eq.~(\ref{Eq:Proof1}) becomes
\begin{align}
    \nonumber
    \hat{\vb{p}}_\xi\delta(\vb{r}-\hat{\vb{r}}_\xi)
    \nonumber
    &=
    \frac{1}{(2\pi\hbar)^3}
    \int\dd[3]{\vb{p}}
    e^{i\vb{p}\cdot(\vb{r}-\hat{\vb{r}}_\xi)/\hbar}\hat{\vb{p}}_\xi\\
    &~-
    \frac{1}{(2\pi\hbar)^3}
    \int\dd[3]{\vb{p}}
    \vb{p}
    e^{i\vb{p}\cdot(\vb{r}-\hat{\vb{r}}_\xi)/\hbar}.
    \label{Eq:Proof3}
\end{align}
In Eq.~(\ref{Eq:Proof3}), we convert the infinite series back to exponential functions.
Note that the integrand of the second term in Eq.~(\ref{Eq:Proof3}) can be further rewritten by
\begin{align}
    -\vb{p}
    e^{i\vb{p}\cdot(\vb{r}-\hat{\vb{r}}_\xi)/\hbar}=
    i\hbar\nabla
    e^{i\vb{p}\cdot(\vb{r}-\hat{\vb{r}}_\xi)/\hbar},
\end{align}
where the gradient operator acts on functions containing $\vb{r}$.
Taking the momentum integral, we obtain the identity
\begin{align}
    \hat{\vb{p}}_\xi\delta(\vb{r}-\hat{\vb{r}}_\xi)=
    \delta(\vb{r}-\hat{\vb{r}}_\xi)\hat{\vb{p}}_\xi+
    i\hbar\nabla\delta(\vb{r}-\hat{\vb{r}}_\xi).
\end{align}
Finally, it is straightforward to obtain that
\begin{align}
    \nonumber
    &\sum_{\xi\in M}\frac{q_\xi}{2m_\xi}\int \dd[3]{\vb{r}}
    \hat{\mathbf{A}}(\mathbf{r})\cdot
    \hat{\vb{p}}_\xi\delta(\vb{r}-\hat{\vb{r}}_\xi)\\
    \nonumber
    &=
    \sum_{\xi\in M}\frac{q_\xi}{2m_\xi} \int \dd[3]{\vb{r}}
    \hat{\mathbf{A}}(\mathbf{r})\cdot
    \delta(\vb{r}-\hat{\vb{r}}_\xi)\hat{\vb{p}}_\xi\\
    &~+
    \sum_{\xi\in M}\frac{q_\xi}{2m_\xi} \int \dd[3]{\vb{r}}
    (-i\hbar)\delta(\vb{r}-\hat{\vb{r}}_\xi)
    \nabla\cdot\hat{\mathbf{A}}(\mathbf{r})
    \label{Eq:Proof4}
    \\
    &=
    \sum_{\xi\in M}\frac{q_\xi}{2m_\xi} \int \dd[3]{\vb{r}}
    \hat{\mathbf{A}}(\mathbf{r})\cdot
    \delta(\vb{r}-\hat{\vb{r}}_\xi)\hat{\vb{p}}_\xi,
\end{align}
where we utilize integration by parts to get 
$\nabla\cdot\hat{\mathbf{A}}(\mathbf{r})$.
Moreover, the choosing of the Coulomb gauge makes $\nabla\cdot\hat{\mathbf{A}}(\mathbf{r})=0$. Thus, the order of $\delta(\vb{r}-\hat{\vb{r}}_\xi)$ and $\hat{\vb{p}}_\xi$ does not alter the result in the Coulomb gauge.

\section{Derivation of Eq.~(\ref{Eq:HABint})}
\label{sect:coupling}
In this appendix, we show the detail derivation of Eq.~(\ref{Eq:HABint}).
According to the definition in Eq.~(\ref{Eq:Hintdef}), we divide the discussion into two parts, the scalar coupling [$\hat{\varphi}(\mathbf{r}){\hat{\rho}}_M(\mathbf{r})$] and the vector coupling [$\hat{\vb{A}}(\vb{r})\cdot\hat{\vb{j}}_M(\vb{r})$]. The key step to derive Eq.~(\ref{Eq:HABint}) is to express the scalar coupling in terms of a current-like operator.
Because the scalar coupling is associated with the longitudinal interaction, we begin with the following quantity,
\begin{align}
    \nonumber
    &\int \dd[3]{\vb{r}}
    \bra{\left\{0\right\}}\hat{\vb{E}}(\mathbf{r})\ket{\left\{1_l(\vb{s,\omega})\right\}}
    \cdot{\vb{j}}_{mm'(m'm)}^\parallel(\mathbf{r})\\
    \nonumber
    &=\int \dd[3]{\vb{r}}
    \bra{\left\{0\right\}}\hat{\vb{E}}^\parallel(\mathbf{r})\ket{\left\{1_l(\vb{s,\omega})\right\}}
    \cdot{\vb{j}}_{mm'(m'm)}(\mathbf{r})\\
    \nonumber
    &=
    \int \dd[3]{\vb{r}}
    \bra{\left\{0\right\}}\hat{\varphi}(\mathbf{r})\ket{\left\{1_l(\vb{s},\omega)\right\}}
    \Big[\nabla\cdot
    {\vb{j}}_{mm'(m'm)}(\mathbf{r})\Big]\\
    \label{Eq:subst_continuityEq}
    &=-
    \int \dd[3]{\vb{r}}
    \bra{\left\{0\right\}}\hat{\varphi}(\mathbf{r})\ket{\left\{1_l(\vb{s},\omega)\right\}}
    {\dot{\rho}}_{mm'(m'm)}(\mathbf{r}),
\end{align}
where 
\begin{subequations}
    \begin{align}
        {\vphantom{j}\vb{j}}_{mm'(m'm)}(\vb{r})&\equiv\bra{m(m')}{\hat{\vb{j}}}_M(\vb{r})\ket{m'(m)},\\
        {\vphantom{\rho}\rho}_{mm'(m'm)}(\mathbf{r})&\equiv\bra{m(m')}{\hat{\rho}}_M(\vb{r})\ket{m'(m)},
    \end{align}
\end{subequations}
are the elements of transition charge density and transition current density for the entity $M$.
To obtain Eq.~(\ref{Eq:subst_continuityEq}), we first use the symmetry of orthogonal projections to get the longitudinal electric field, and then take the integration by parts to get the divergence of the transition current density.
Furthermore, by utilizing the continuity equation, $\nabla\cdot{\vb{j}}_{mm'(m'm)}(\vb{r})=-{\dot{\rho}}_{mm'(m'm)}(\vb{r})$, we replace the divergence of the transition current density by the time derivative of transition charge density.
It is worth noting that the continuity equation is valid as we suppose that charge transfer processes do not happen between entities, i.e., the number of charges for each entity is conserved.
In addition, the time derivative transition charge density ${\dot{\rho}}_{mm'(m'm)}(\mathbf{r})$ can be replaced according to the relation,
\begin{align}
    \nonumber
    {\dot{\rho}}_{mm'(m'm)}(\vb{r})
    =&
    \bra{m(m')}
    \frac{1}{i\hbar}\comm{{\hat{\rho}}_M(\vb{r})}{\hat{H}_M}
    \ket{m'(m)}\\
    =&
    \mp i\omega_{m'm}~{\rho}_{mm'(m'm)}(\vb{r}),
    \label{Eq:derrho}
\end{align}
where $\omega_{m'm}=\omega_{m'}-\omega_{m}$ and $\hbar\omega_{m'(m)}$ is the energy of the state $\ket{m'(m)}$.
Substituting Eq.~(\ref{Eq:derrho}) into Eq.~(\ref{Eq:subst_continuityEq}), we obtain the relation,
\begin{align}
    \nonumber
    &\int \dd[3]{\vb{r}}
    \bra{\left\{0\right\}}\hat{\varphi}(\mathbf{r})\ket{\left\{1_l(\vb{s},\omega)\right\}}\rho_{mm'(m'm)}(\vb{r})\\
    &=\int \dd[3]{\vb{r}}
    \bra{\left\{0\right\}}\hat{\vb{E}}(\mathbf{r})\ket{\left\{1_l(\vb{s,\omega})\right\}}\cdot
    \frac{{\vb{j}}_{mm'(m'm)}^\parallel(\vb{r})}{\pm i\omega_{m'm}}.
    \label{Eq:scalarcouplingrelation}
\end{align}
Moreover, it is convenient to express the couplings in the frequency domain because the polariton states are portrayed in the frequency domain.
By definition, the electric-field operator is the integral of the auxiliary electric-field operator [defined in Eq.~(\ref{Eq:Edef})] and its Hermitian conjugate,
\begin{align}
    \hat{\vb{E}}(\vb{r}) = \int_0^\infty \dd{\omega}
    \Big[\hat{\vb{E}}(\vb{r},\omega) +\hc\Big].
    \label{Eq:longEfreq}
\end{align}
By inserting Eq.~(\ref{Eq:longEfreq}) into Eq.~(\ref{Eq:scalarcouplingrelation}) and defining the longitudinal auxiliary transition current density as,
\begin{subequations}
\begin{align}
    {\mathcal{J}}_{mm'}^\parallel(\vb{r};-\omega)&\equiv
    \frac{-\omega}{\omega_{m'm}}
    {\vb{j}}_{mm'}^\parallel(\vb{r}),\\
    {\mathcal{J}}_{m'm}^\parallel(\vb{r};+\omega)&\equiv
    \frac{+\omega}{\omega_{m'm}}
    {\vb{j}}_{m'm}^\parallel(\vb{r}).
\end{align}
\label{Eq:auxiliaryJlong}%
\end{subequations}%
we obtain that 
\begin{widetext}

\begin{subequations}
\begin{align}
    &\int \dd[3]{\vb{r}}
    \bra{\left\{0\right\}}\hat{\varphi}(\mathbf{r})\ket{\left\{1_l(\vb{s},\omega)\right\}}
    {\rho}_{mm'}(\mathbf{r})
    =-\int \dd[3]{\vb{r}} \int \dd{\omega'}
    (i\omega')^{-1}
    \bra{\left\{0\right\}}\hat{\vb{E}}(\mathbf{r},\omega')\ket{\left\{1_l(\vb{s,\omega})\right\}}
    \cdot{\mathcal{J}}_{mm'}^\parallel(\vb{r};-\omega'),\\
    &\int \dd[3]{\vb{r}} 
    \bra{\left\{0\right\}}\hat{\varphi}(\mathbf{r})\ket{\left\{1_l(\vb{s},\omega)\right\}}
    {\rho}_{m'm}(\mathbf{r})
    =-\int \dd[3]{\vb{r}} \int \dd{\omega'}
    (i\omega')^{-1}
    \bra{\left\{0\right\}}\hat{\vb{E}}(\mathbf{r},\omega')\ket{\left\{1_l(\vb{s,\omega})\right\}}
    \cdot{\mathcal{J}}_{m'm}^\parallel(\vb{r},\omega').
\end{align}
\label{Eq:HintABlong}
\end{subequations}

\noindent
Note that the denominator $\omega_{m'm}$ in Eq.~(\ref{Eq:auxiliaryJlong}) is always defined as the positive transition frequency whatever the inital and final eigenkets are.
Note that the electric field operator in the frequency domain contains $\hat{\vb{E}}(\vb{r},\omega)$ and its Hermitian conjugate $\hat{\vb{E}}^\dagger(\vb{r},\omega)$. Here, only the operator $\hat{\vb{E}}(\vb{r},\omega)$ is preserved because the element of $\hat{\vb{E}}^\dagger(\vb{r},\omega)$ gives a zero, i.e.,
\begin{align}
    &\bra{\left\{0\right\}}\hat{\vb{E}}^\dagger(\mathbf{r},\omega')\ket{\left\{1_l(\vb{s},\omega)\right\}}
    \propto\bra{\left\{0\right\}}\hat{\vb{f}}^\dagger(\mathbf{r},\omega')\ket{\left\{1_l(\vb{s},\omega)\right\}}
    =0.
\end{align}

Similarly, by using the definition of the vector potential operator in Eq.~(\ref{Eq:defA}), we rewrite the vector coupling [$\hat{\vb{A}}(\vb{r})\cdot\hat{\vb{j}}_M(\vb{r})$] in the frequency domain, 
\begin{subequations}
\begin{align}
    &\int\dd[3]{\vb{r}}
    \bra{\left\{0\right\}}\hat{\mathbf{A}}(\mathbf{r})\ket{\left\{1_l(\vb{s},\omega)\right\}}
    \cdot\vb{j}_{mm'}(\vb{r})
    =
    \int\dd[3]{\vb{r}}\int_0^\infty \dd{\omega'}
    (i\omega')^{-1}
    \bra{\left\{0\right\}}
    \hat{\mathbf{E}}(\mathbf{r},\omega')\ket{\left\{1_l(\vb{s},\omega)\right\}}
    \cdot\mathcal{J}_{mm'}^\perp(\vb{r};-\omega'),\\
    &\int\dd[3]{\vb{r}}
    \bra{\left\{0\right\}}\hat{\mathbf{A}}(\mathbf{r})\ket{\left\{1_l(\vb{s},\omega)\right\}}
    \cdot\vb{j}_{m'm}(\vb{r})
    =
    \int\dd[3]{\vb{r}}\int_0^\infty \dd{\omega'}
    (i\omega')^{-1}
    \bra{\left\{0\right\}}
    \hat{\mathbf{E}}(\mathbf{r},\omega')\ket{\left\{1_l(\vb{s},\omega)\right\}}
    \cdot\mathcal{J}_{m'm}^\perp(\vb{r};\omega'),
\end{align}%
\label{Eq:HintABtrans}%
\end{subequations}%
%
where we use
\begin{align}
    {\vb{j}}_{mm'(m'm)}^\perp(\vb{r})={\mathcal{J}}_{mm'(m'm)}^\perp(\vb{r};\omega),
    \label{Eq:auxiliaryJtrans}
\end{align}
to denote the transverse auxiliary transition current density because the transverse part of the auxiliary transition current density is independent of $\omega$.
From Eqs.~(\ref{Eq:auxiliaryJlong}) and (\ref{Eq:auxiliaryJtrans}), the total auxiliary transition current density finally becomes
\begin{align}
    &{\mathcal{J}}_{mm'(m'm)}(\vb{r};\omega)
    =
    {\mathcal{J}}_{mm'(m'm)}^\parallel(\vb{r};\omega)+
    {\mathcal{J}}_{mm'(m'm)}^\perp(\vb{r};\omega).
    \label{Eq:AuxiliaryJ}
\end{align}
According to Eqs.~(\ref{Eq:HintABlong}), (\ref{Eq:HintABtrans}), (\ref{Eq:AuxiliaryJ}), and (\ref{Eq:Hintdef}), the above elements of $\hat{V}_{\mathrm{pol},M}$ become
\begin{subequations}
\begin{align}
        &\bra{a;\left\{0\right\}}
        \hat{V}_{\mathrm{pol},A}
        \ket{a';\left\{1_l(\vb{s},\omega)\right\}}=
        -\int\dd[3]{\vb{r}}\int_0^\infty \dd{\omega'}
        (i\omega')^{-1}
        \bra{\left\{0\right\}}\hat{\mathbf{E}}(\mathbf{r},\omega')\ket{\left\{1_l(\vb{s},\omega)\right\}}\cdot{\mathcal{J}}_{aa'}(\vb{r};-\omega'),\\
        &\bra{b';\left\{0\right\}}
        \hat{V}_{\mathrm{pol},B}
        \ket{b;\left\{1_l(\vb{s},\omega)\right\}}=
        -\int\dd[3]{\vb{r}}\int_0^\infty \dd{\omega'}
        (i\omega')^{-1}
        \bra{\left\{0\right\}}\hat{\mathbf{E}}(\mathbf{r},\omega'
        )\ket{\left\{1_l(\vb{s},\omega)\right\}}\cdot{\mathcal{J}}_{b'b}(\vb{r};\omega').
\end{align}%
\label{Eq:HABintrest1}
\end{subequations}
Similarly, by taking the same procedure addressed above, one can obtain the rest of the elements listed in Eq.~(\ref{Eq:HABint}),
\begin{subequations}
\begin{align}
    &\bra{b';\left\{1_l(\vb{s},\omega)\right\}}
    \hat{V}_{\mathrm{pol},B}
    \ket{b;\left\{0\right\}}=
    \int\dd[3]{\vb{r}}\int_0^\infty \dd{\omega'}
    (i\omega')^{-1}
    \bra{\left\{1_l(\vb{s},\omega)\right\}}\hat{\mathbf{E}}^\dagger(\mathbf{r},\omega')\ket{\left\{0\right\}}\cdot{\mathcal{J}}_{b'b}(\vb{r};-\omega'),\\
    &\bra{a;\left\{1_l(\vb{s},\omega)\right\}}
    \hat{V}_{\mathrm{pol},A}
    \ket{a';\left\{0\right\}}=
    \int\dd[3]{\vb{r}}\int_0^\infty \dd{\omega'}
    (i\omega')^{-1}
    \bra{\left\{1_l(\vb{s},\omega)\right\}}\hat{\mathbf{E}}^\dagger(\mathbf{r},\omega')\ket{\left\{0\right\}}\cdot{\mathcal{J}}_{aa'}(\vb{r};\omega').
\end{align}%
\label{Eq:HABintrest2}
\end{subequations}
Incidentally, it is easy to check the correctness of Eq.~(\ref{Eq:HABintrest2}) by taking the complex conjugate of Eq.~(\ref{Eq:HABintrest1}).
\end{widetext}

\section{$\omega$-Integral Along the Path $C_2$}
\label{sect:Gamma2}
In this appendix, we present the details of evaluating the line integral of $I(\vb{r},\vb{r}',\omega)$ along the semicircular path $C_2$ in the upper half-plane.
By definition, the integral is equal to 
\begin{align}
    \int_{C_2} I(\vb{r},\vb{r}',\Omega) \dd{\Omega}
    &= \lim_{\abs{\Omega}\rightarrow\infty}\int_0^\pi I(\vb{r},\vb{r}',\Omega)~i\Omega \dd{\theta}.
    \label{Eq:intGamma2}
\end{align}
Recall that $\Omega=\omega e^{i\theta}$.
Here, we define $I(\vb{r},\vb{r}',\Omega)=I_1(\vb{r},\vb{r}',\Omega)+I_2(\vb{r},\vb{r}',\Omega)$, where the two functions are
\begin{align*}
    I_1(\vb{r},\vb{r}',\Omega) &= 
    \frac{1}{2ic^2}\frac{
    {\mathcal{J}}_{b'b}(\vb{r};\Omega)
    \cdot \tensor{\vb{G}}(\vb{r},\vb{r}',\Omega)\cdot
    {\mathcal{J}}_{aa'}(\vb{r}';\Omega)}
    {\hbar(\omega_{\mathrm{T}}-\Omega)+i\eta\sgn{\Omega}},\\
    I_2(\vb{r},\vb{r}',\Omega) &= 
    \frac{-1}{2ic^2}\frac{
    {\mathcal{J}}_{b'b}(\vb{r};\Omega)
    \cdot \tensor{\vb{G}}\vphantom{G}^*(\vb{r},\vb{r}',\Omega)\cdot
    {\mathcal{J}}_{aa'}(\vb{r}';\Omega)}
    {\hbar(\omega_{\mathrm{T}}-\Omega)+i\eta\sgn{\Omega}},
\end{align*}
respectively. Using the identity of the dyadic Green's function,
\begin{align}
    \lim_{\abs{\Omega}\rightarrow\infty}\frac{\Omega^2}{c^2}\tensor{\vb{G}}(\mathbf{r},\mathbf{r'},\Omega) 
    = -\tensor{\vb{I}}\delta(\mathbf{r-r'}),
    \label{Eq:dyadGlim}
\end{align}
we evaluate the limit of $i\Omega~I_1(\vb{r},\vb{r}',\Omega)$ and obtain that
\begin{align}
    \lim_{\abs{\Omega}\rightarrow\infty}i\Omega~I_1(\vb{r},\vb{r}',\Omega)=
    \frac{1}{2\omega_\mathrm{T}^2}
    {\vb{j}}_{b'b}^\parallel(\vb{r})\cdot
    {\vb{j}}_{aa'}^\parallel(\vb{r}')~\delta(\vb{r}-\vb{r}').
    \label{Eq:I1lim}
\end{align}
Note that the asymptotic behavior of the auxiliary current density is 
\begin{align}
    \nonumber
    {\mathcal{J}}_{m'm(mm')}^\parallel(\vb{r};\Omega)\sim
    \frac{\Omega}{\omega_{m'm}}~
    {\vb{j}}_{m'm(mm')}^\parallel(\vb{r}), \quad
    \abs{\Omega}\rightarrow\infty,
\end{align}
and recall that $\omega_{a'a}=\omega_{b'b}=\omega_\mathrm{T}$.
The similar procedure is done on $I_2(\vb{r},\vb{r}',\omega)$, but we should utilize the Schwarz reflection principle of the dyadic Green's function~\cite{Buhmann2012} and transform the integral variable to $\Omega\rightarrow-\Omega^*$ first.
Finally, we obtain the same result,
\begin{align}
    \lim_{\abs{\Omega}\rightarrow\infty}i\Omega~I_2(\vb{r},\vb{r}',\Omega)=
    \frac{1}{2\omega_\mathrm{T}^2}
    {\vb{j}}_{b'b}^\parallel(\vb{r})\cdot
    {\vb{j}}_{aa'}^\parallel(\vb{r}')~\delta(\vb{r}-\vb{r}').
    \label{Eq:I2lim}
\end{align}
It is obvious that the integrand in Eq.~(\ref{Eq:intGamma2}) is no longer dependent of $\theta$ after taking the limit, as shown in Eqs.~(\ref{Eq:I1lim}) and (\ref{Eq:I2lim}). Thus, the integral associated with $I_1(\vb{r},\vb{r}',\omega)$ is $\pi$.
Moreover, because we take the transformation $\Omega\rightarrow-\Omega^*$, the integral becomes
\begin{align}
    \int_0^\pi \dd{\theta}\rightarrow-\int_0^{-\pi} \dd{\theta}= \pi.
    \label{Eq:I2int}
\end{align}
According to Eqs.~(\ref{Eq:I1lim}),~(\ref{Eq:I2lim}), and~(\ref{Eq:I2int}), Eq.~(\ref{Eq:intGamma2}) eventually becomes
\begin{align}
    \int_{C_2} I(\vb{r},\vb{r}',\Omega) \dd{\Omega}
    &= \frac{\pi}{\hbar\omega_\mathrm{T}^2}
    {\vb{j}}_{b'b}^\parallel(\vb{r})\cdot
    {\vb{j}}_{aa'}^{\parallel}(\vb{r}')
    ~\delta(\vb{r}-\vb{r}').
    \label{Eq:Gamma2_3}
\end{align}

\section{Interaction of Longitudinal Transition Current Densities}
\label{sect:longcurrent}
We prove that the interaction between two longitudinal transition current densities is exactly the Coulomb interaction, namely,
\begin{align}
    \label{Eq:jlongAppendD}
    \frac{1}{\epsilon_0\omega_\mathrm{T}^2}
    \int\dd[3]{\vb{r}}~
    {\vb{j}}_{b'b}^\parallel(\vb{r})\cdot
    {\vb{j}}_{aa'}^\parallel(\vb{r})
    &=
    \bra{a;b'}
    \hat{V}_{AB}
    \ket{a';b}.
\end{align}
According to the definition of longitudinal fields~\cite{jackson1998classical}, the longitudinal transition current density can be described as follows,
\begin{align}
    \nonumber
    {\vb{j}}_{m'm(mm')}^\parallel(\vb{r}) 
    &\equiv \frac{-1}{4\pi}\nabla \int \dd[3]{\vb{r}'} \frac{\nabla'\cdot{\vb{j}}_{m'm(mm')}(\vb{r}')}
    {\abs{\vb{r}-\vb{r}'}}\\
    &=\frac{\pm 1}{4\pi}\nabla \int \dd[3]{\vb{r}'} \frac{i\omega_\mathrm{T}~{\rho}_{m'm(mm')}(\vb{r}')}{\abs{\vb{r}-\vb{r}'}}.
\end{align}
Note that $\nabla\cdot{\vb{j}}_{m'm}(\vb{r})=- i\omega_\mathrm{T}\rho_{m'm}(\vb{r})$ and
$\nabla\cdot{\vb{j}}_{mm'}(\vb{r})= i\omega_\mathrm{T}\rho_{mm'}(\vb{r})$
[Recall Eq.~(\ref{Eq:derrho})]. By using the definition of charge density operator [Eq.~(\ref{Eq:Mrhodef})], the longitudinal transition current density of the molecule $M$ becomes
\begin{align}
    \nonumber
    &{\vb{j}}_{m'm(mm')}^\parallel(\vb{r})\\
    &=\pm\frac{i\omega_\mathrm{T}}{4\pi}\nabla
    \bra{m'(m)}\sum_{\xi\in M}\frac{q_\xi}{\abs{\vb{r}-\hat{\vb{r}}_\xi}}
    \ket{m(m')}.
\end{align}
Thus, the left-hand side (LHS) of Eq.~(\ref{Eq:jlongAppendD}) becomes
\begin{align}
    \nonumber
    \mathrm{LHS}&
    =\frac{1}{16\pi^2\epsilon_0}\int\dd[3]{\vb{r}}\\
    &~\times
    \bra{a;b'}
    \sum_{\xi\in A}\sum_{\zeta\in B} q_\xi q_{\zeta}
    K(\vb{r},\hat{\vb{r}}_\xi,\hat{\vb{r}}_{\zeta})
    \ket{a';b},
    \label{Eq:LHS_T2gamma2}
\end{align}
where
\begin{align}
    K(\vb{r},\hat{\vb{r}}_\xi,\hat{\vb{r}}_{\zeta}) 
    &=
    \nabla g(\vb{r},\hat{\vb{r}}_\xi)\cdot
    \nabla g(\vb{r},\hat{\vb{r}}_{\zeta}),
\end{align}
with $g(\vb{r},\hat{\vb{r}}_\xi)=1/\abs{\vb{r}-\hat{\vb{r}}_\xi}$ being the Green's function of Poisson's equation,
\begin{align}
    \nabla^2 g(\vb{r},\hat{\vb{r}}_\xi) = -4\pi\delta(\vb{r}-\hat{\vb{r}}_\xi).
    \label{Eq:defGreenFunc_Poisson}
\end{align}
By using the Green's identity, $K(\vb{r},\hat{\vb{r}}_\alpha,\hat{\vb{r}}_\beta)$ becomes
\begin{align}
    \nonumber
    &K(\vb{r},\hat{\vb{r}}_\xi,\hat{\vb{r}}_{\zeta})\\
    &=
    \nabla \cdot \Big[ g(\vb{r},\hat{\vb{r}}_\xi) \cdot
    \nabla g(\vb{r},\hat{\vb{r}}_{\zeta}) \Big]-
    g(\vb{r},\hat{\vb{r}}_\xi)
    \nabla^2 g(\vb{r},\hat{\vb{r}}_{\zeta}).
    \label{Eq:K_GreenIdentity}
\end{align}
Plugging the result of $K(\vb{r},\hat{\vb{r}}_\xi,\hat{\vb{r}}_{\zeta})$ into Eq.~(\ref{Eq:LHS_T2gamma2}) and using the definition of the Green's function in Eq.~(\ref{Eq:defGreenFunc_Poisson}), we obtain that
\begin{align}
    \nonumber
    \mathrm{LHS}&=\frac{1}{4\pi\epsilon_0}\bra{a;b'}
    \sum_{\xi\in A}\sum_{\zeta\in B} q_\xi q_{\zeta}
    g(\hat{\vb{r}}_\xi,\hat{\vb{r}}_{\zeta})
    \ket{a';b}\\
    &=\bra{a;b'}\hat{V}_{AB}
    \ket{a';b},
\end{align}
which is exactly the Coulomb interaction between molecules $A$ and $B$.
Note that the first term in Eq.~(\ref{Eq:K_GreenIdentity}) associated with a surface integral at $\vb{r}\rightarrow\infty$, thus it converges to zero after the integration.

\section{Transition Current Density and Single-Electron Wavefunction}
\label{sect:TCD}
In the appendix, we show that the transition current density can be expressed by single-electron wavefunctions.
Beginning from the consequence of the Condon-like approximation in Eq.~(\ref{Eq:CondonlikeApprox}), we separate the nuclear degrees of freedom and define the electronic transition current density as follows,
\begin{align}
    \nonumber
    &\vb{j}_{\gamma\gamma'}^M(\vb{r})\\
    \nonumber
    &\equiv\matrixel*{\phi_{M,\gamma}}{\hat{\vb{j}}_M(\vb{r})}{\phi_{M,\gamma'}}_{\{\vb{R}\}}\\
    &=
    \bra{\phi_{M,\gamma}}
    \sum_{\xi\in M_\mathrm{el}}\frac{q_\xi}{2m_\xi}\left[
    \delta(\vb{r}-\hat{\vb{r}}_\xi)\hat{\vb{p}}_\xi+\hc\right]\ket{\phi_{M,\gamma'}}_{\{\vb{R}\}},
    \label{Eq:j_edef}
\end{align}
where $M_\mathrm{el}$ is the set of electrons in the entity $M$.
To project the states to the position-spin coordinates, we use the fermionic completeness relation~\cite{Gianluca2013},
\begin{align}
    \frac{1}{N!}\int \mathcal{D}\{\vb{x}_\mu\} 
    \dyad{\{\vb{x}_\mu\}}{\{\vb{x}_\mu\}}= \hat{\vb{I}},
\end{align}
where $N$ is the total number of position-spin coordinates, $\hat{\vb{I}}$ is the identity operator, and the integration symbol is defined as
\begin{align}
    \int \mathcal{D}\{\vb{x}_\mu\} \equiv \int \dd{\vb{x}_1} \dots \int \dd{\vb{x}_N}.
\end{align}
Here, we denotes the $\mu$-th position-spin coordinate as $\vb{x}_\mu \equiv (\vb{r}_\mu,\omega_\mu)$, where $\omega_\mu$ is the spin coordinate.
Moreover, the antisymmetrized state $\ket{\{\vb{x}_\mu\}}\equiv\ket{\vb{x}_1\dots\vb{x}_N}$ is the collection of position-spin coordinates.
After inserting the identity to Eq.~(\ref{Eq:j_edef}), the electronic transition current density becomes
\begin{widetext}
\begin{align}
    \nonumber
    \vb{j}_{\gamma\gamma'}^M(\vb{r})
    &= \frac{1}{N_\mathrm{el}!^3}
    \int \mathcal{D}\{\vb{x}_\mu\} \int \mathcal{D}\{\vb{x}'_\mu\}
    \int \mathcal{D}\{\vb{x}''_\mu\}\\
    &~\times
    \sum_{\xi\in M_\mathrm{el}}\frac{q_\xi}{2m_\xi}
    \braket{\phi_{M,\gamma}}{\{\vb{x}_\mu\}}_{\{\vb{R}\}}
    \bigg[\bra{\{\vb{x}_\mu\}}
    \delta(\vb{r}-\hat{\vb{r}}_\xi)\dyad*{\{\vb{x}_\mu'\}}
    \hat{\vb{p}}_\xi\ket*{\{\vb{x}_\mu''\}}
    + \mathrm{c.c.}\bigg]
    \braket{\{\vb{x}_\mu''\}}{\phi_{M,\gamma'}}_{\{\vb{R}\}}.
    \label{Eq:jexpansion}
\end{align}

\noindent
Recall that $N_\mathrm{el}$ is the total number of electrons in the entity $M$, and the subscript $\{\vb{R}\}$ represents the electronic state is based on specific nuclear coordinates.
In Eq.~(\ref{Eq:jexpansion}), the projection of states to position-spin coordinates is related to the multi-electron wavefunction. According to the previous study~\cite{Gianluca2013}, the multi-electron wavefunction can be expressed as
\begin{align}
    \phi_{M,\gamma(\gamma')}(\{\vb{x}_\mu\};{\{\vb{R}\}})=
    \frac{1}{\sqrt{N!}}\braket{\{\vb{x}_\mu\}}{\phi_{M,\gamma(\gamma')}}_{\{\vb{R}\}}.
\end{align}
Note that $\phi_{M,\gamma(\gamma')}(\{\vb{x}_\mu\})$ is normalized. Furthermore,
according to the following two identities~\cite{Gianluca2013},
\begin{align}
    &\bra{\{\vb{x}_\mu\}}
    \delta(\vb{r}-\hat{\vb{r}}_\xi)\ket*{\{\vb{x}_\mu'\}}
    =
    \delta(\vb{r}-\vb{r}_\xi)\sum_P(-1)^P\prod_{\mu\in M_\mathrm{el}}\delta\Big(\vb{x}_\mu-\vb{x}_{P(\mu)}'\Big),
    \label{Eq:detaelement}
\end{align}
and 
\begin{align}
    &\bra*{\{\vb{x}_\mu'\}}
    \hat{\vb{p}}_\xi\ket*{\{\vb{x}_\mu''\}}
    =
    -i\hbar\nabla_\xi'\sum_P(-1)^P\prod_{\mu\in M_\mathrm{el}}\delta\Big(\vb{x}_\mu'-\vb{x}_{P(\mu)}''\Big),
    \label{Eq:pelement}
\end{align}
we can simplify the electronic transition current density to
\begin{align}
    \vb{j}_{\gamma\gamma'}^M(\vb{r})
    \nonumber
    =-\frac{i\hbar eN_\mathrm{el}}{2m_\mathrm{el}}
    \int \mathcal{D}\{\vb{x}'_\mu\}~\delta(\vb{r}-\vb{r}_1)
    \Big[
    &\phi_{M,\gamma}^*(\{\vb{x}_\mu'\};{\{\vb{R}\}})~\nabla_1'~
    \phi_{M,\gamma'}(\{\vb{x}_\mu'\};{\{\vb{R}\}})\\
    -&\phi_{M,\gamma'}(\{\vb{x}_\mu'\};{\{\vb{R}\}})~\nabla_1'~
    \phi_{M,\gamma}^*(\{\vb{x}_\mu'\};{\{\vb{R}\}})
    \Big],
    \label{Eq:j1integral}
\end{align}
where $e$ is the elementary charge and $m_\mathrm{el}$ is electron mass.
Note that the summations of $P$ in Eqs.~(\ref{Eq:detaelement}) and (\ref{Eq:pelement}) denote the summations of all different permutations.
To obtain Eq.~(\ref{Eq:j1integral}), we simplify the summation of $\xi\in M_\mathrm{el}$ by multiplying $N_\mathrm{el}$ because electrons are indistinguishable, i.e., the total electronic transition current density is $N_\mathrm{el}$-times of the single-electron transition current density.
Therefore, we express the single-electron transition current density by choosing $\xi=1$ ($\vb{r}_\xi = \vb{r}_1$ and $\nabla_\xi'=\nabla_1'$).
The other electronic degrees of freedoms for $\mu\neq\xi$ in multi-electron wavefunctions ($\phi_{M,\gamma'}$ and $\phi_{M,\gamma}^*$) have been taken account into the overlap integral in Eq.~(\ref{Eq:eff1ewfn}),
\begin{align}
    \nonumber
    \tilde{\phi}_{M,\gamma}^*(\vb{x}_1;\{\vb{R}\})\tilde{\phi}_{M,\gamma'}(\vb{x}_1;\{\vb{R}\})\equiv
    \int \mathcal{D}\{\vb{x}_{\mu\neq1}\}~\phi_{M,\gamma}^*(\{\vb{x}_\mu\};\{\vb{R}\})\phi_{M,\gamma'}(\{\vb{x}_\mu\};\{\vb{R}\}),
\end{align}
and finally we derive the electronic transition current density as follows,

\begin{align}
    \vb{j}_{\gamma\gamma'}^M(\vb{r})&=
    \frac{-i\hbar e N_\mathrm{el}}{2m_\mathrm{el}}
    \bigg[
    \tilde{\phi}_{M,\gamma}^*(\vb{r};\{\vb{R}\})\nabla\tilde{\phi}_{M,\gamma'}(\vb{r};\{\vb{R}\})
    -
    \tilde{\phi}_{M,\gamma'}(\vb{r};\{\vb{R}\})\nabla\tilde{\phi}_{M,\gamma}^*(\vb{r};\{\vb{R}\})
    \bigg].
\end{align}

\end{widetext}

\nocite{*}


%

\end{document}